\documentclass[fullpage]{article}
\newcounter{mnotecount}[section]

\renewcommand{\themnotecount}{\thesection.\arabic{mnotecount}}

\newcommand{\mnote}[1]
{\protect{\stepcounter{mnotecount}}$^{\mbox{\footnotesize
$
\bullet$\themnotecount}}$ \marginpar{
\raggedright\tiny\em
$\!\!\!\!\!\!\,\bullet$\themnotecount: #1} }

\usepackage[utf8]{inputenc}
\usepackage{amsmath,amssymb}
\usepackage{amsmath}
\usepackage{bbm}
\usepackage{amsthm}
\usepackage{enumerate}
\usepackage{braket}
\usepackage{bm}
\usepackage[title]{appendix}
\usepackage{tikz}
\usepackage[none]{hyphenat}
\usepackage{geometry}
\usepackage{float}
\usepackage{fancyhdr}
\usepackage{lipsum}
\usepackage{braket}
\usepackage[labelfont=bf]{caption}
\theoremstyle{definition}

\usepackage{pifont}
\newcommand{\cmark}{\ding{51}}%
\newcommand{\xmark}{\ding{55}}%

\title{On Schwarzschild causality in higher dimensions}
\author{Peter Cameron\footnote{pjc96@cam.ac.uk}\;\; and Maciej Dunajski\footnote{m.dunajski@damtp.cam.ac.uk}\\
Department of Applied Mathematics and Theoretical Physics\\ 
University of Cambridge\\ Wilberforce Road, Cambridge CB3 0WA, UK.}
\date{August 11, 2020}
\begin{document}
\theoremstyle{definition}
\newtheorem{definition}{Definition}[section]
\newtheorem{exmp}[definition]{Example}
\newtheorem{prop}[definition]{Proposition}
\newtheorem{lemma}[definition]{Lemma}
\newtheorem{thm}[definition]{Theorem}
\maketitle
\begin{abstract}
\noindent 
We show that the causal properties of asymptotically flat spacetimes depend on their dimensionality: while the time-like future of any point
in the past conformal infinity $\mathcal{I}^-$  contains the whole of the future conformal infinity $\mathcal{I}^+$ 
in $(2+1)$ and $(3+1)$ dimensional Schwarzschild 
spacetimes, this property (which we call the Penrose property) does not hold for  $(d+1)$ dimensional Schwarzschild if $d>3$.
We also show that the Penrose property holds for the Kerr solution in $(3+1)$ dimensions, and discuss the connection with scattering theory
in the presence of positive mass.
\end{abstract} 
\section{Introduction}
\begin{figure}
    \centering
    \includegraphics[scale=0.2]{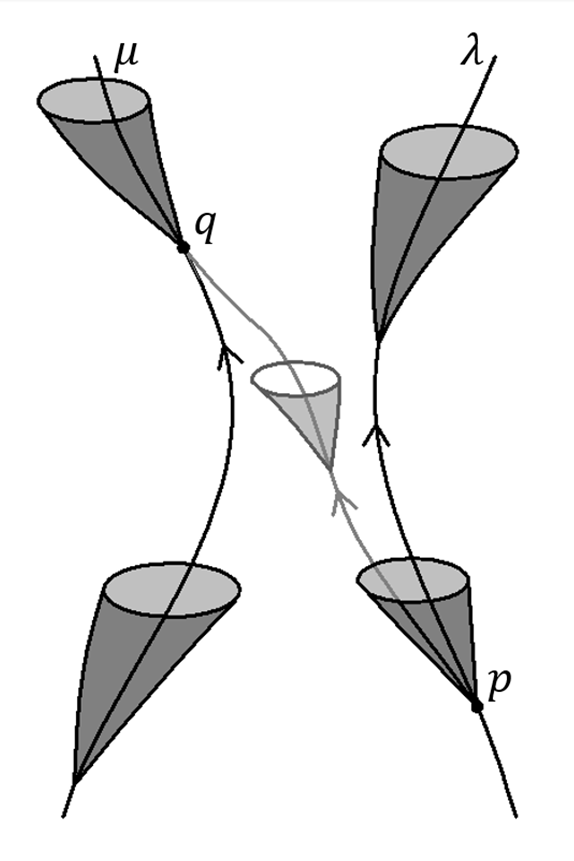}
    \caption{Two endless timelike curves connected by a timelike curve}
    \label{fig:adam}
\end{figure}

In his criticism of a theory of quantum gravity, Penrose \cite{Penrose} has considered the problem of whether any pair of
endless timelike curves in an asymptotically flat Lorentzian spacetime can be connected by a timelike curve (Figure \ref{fig:adam}).
This causal property is conformally invariant, and can be reformulated (Theorem \ref{endless}) in terms of past and future
conformal infinities $\mathcal{I}^-$  and $\mathcal{I}^+$. We shall say that a weakly asymptotically simple Lorentzian manifold (see section \ref{`Lorentz Covariant' Quantum Gravity} for precise assumptions and definitions) admits the Penrose property if and only if the timelike future of any
point $p\in\mathcal{I}^-$ contains the whole of $\mathcal{I}^+$. 
Penrose established that while the property does not hold for Minkowski space (in any dimension), it does in fact hold for positive mass Schwarzschild spacetime in 3+1 dimensions. We will consider the Penrose property in a more general setting and investigate how it is affected by dimension and by the presence of both positive and negative mass. The results can be summarised by the following theorem:\\
\newpage
\textbf{Theorem A:}
{\em
The Penrose property is satisfied by Schwarzschild spacetime of mass $m$ and varying spacetime dimension according to the following table
\begin{center}
    \begin{tabular}{c|c|c}
  Spacetime dimension & $m>0$ & $m\leq0$\\
   \hline
 $3$ & \cmark & \xmark\\ 
 $4$ & \cmark & \xmark\\  
 $\geq5$ & \xmark &  \xmark
    \end{tabular}
\end{center}
}

Besides the quantum gravity motivation of \cite{Penrose}, we also seek to clear up some apparent contradictions in the literature. For example, in a footnote in section 2 of \cite{WittenLightRays}, Witten states that one can leave the dimension arbitrary when studying the causal properties of spacetimes and that restricting to 4 spacetime dimensions `does not introduce any complications'. This is in contradiction with \cite{Maldacena}, where in section 2.1 the authors outline an argument which suggests that the causal properties of higher dimensional asymptotically flat spacetimes are different to those in 4 dimensions. The results of Theorem A agree that the causal structure of spacetime is affected by dimensionality. 

We begin in section \ref{`Lorentz Covariant' Quantum Gravity} by summarising Penrose's motivation for studying the Penrose property. This comes from considering, as he describes it, a `Lorentz covariant' approach to quantum gravity. In section \ref{Minkowski Causality} we consider Minkowski spacetime in more detail and determine which pairs of points on $\mathcal{I}^-$ and $\mathcal{I}^+$ can be connected by timelike curves. In section \ref{Schwarzschild of Dimension 2+1}, we then consider the effect of inserting a point mass in 2+1 dimensions. We will see that only in the case of positive mass is the Penrose property satisfied. In section \ref{Positive Mass Schwarzschild in 3+1 Dimensions} we review Penrose's argument that the Penrose property does hold for Schwarzschild in 3+1 dimensions. By comparison with previous results for Minkowski spacetime, we then show in section \ref{Negative Mass Schwarzschild in 3+1 Dimensions} that this is not true in the negative mass case. In sections \ref{Positive Mass Schwarzschild in Higher Dimensions} and \ref{Negative Mass Schwarzschild in Higher Dimensions} we investigate the higher dimensional Schwarzschild metrics of positive and negative mass respectively. The results are summarised again in section \ref{Summary of the Penrose Property} and we observe that the Penrose property appears to be a property of positive mass spacetimes in low dimensions. In section \ref{Kerr Metric} we show how the methods used in previous sections can be extended to study the Penrose property for the (positive mass) Kerr metric in 3+1 dimensions. This case is more complicated because the metric is no longer spherically symmetric. Nevertheless, we are able to prove the following theorem:\\
\\
\textbf{Theorem B:}
{\em The (positive mass) Kerr spacetime in $3+1$ dimensions satisfies the Penrose property.
}
\\
\\
Finally, in section \ref{uniquecontinuation}, we show that the results found previously are consistent with the recent work of \cite{LinearWaves} on  
unique continuation from null infinity of the linear wave equation in various spacetimes. We also discuss the Penrose property for general static, spherically symmetric spacetimes.
\subsection*{Acknowledgements} We are grateful to Mihalis Dafermos, Greg Galloway, Roger Penrose, Harvey Reall, Paul Tod and
Claude Warnick for useful discussions, and to the anonymous referees for their comments which resulted in 
several clarifications.
The work of M.D. has been partially supported by STFC consolidated grant no. ST/P000681/1. P.C. is grateful to St. John's College, Cambridge for their support through a College Scholarship from the Todd/Goddard Fund. Some of this work was carried out while P.C. was visiting Institut Mittag-Leffler in Djursholm, Sweden, where he was supported by the Swedish Research Council under grant no. 2016-06596.
\section{`Lorentz Covariant' Quantum Gravity}
\label{`Lorentz Covariant' Quantum Gravity}
In \cite{Penrose}, Penrose considers a `Lorentz covariant' approach to quantum gravity where the theory is constructed with respect to some background Minkowski metric $\eta$. The physical field tensors, including the metric $g$, are constructed as an infinite sum of tensors defined with respect to this background metric. For example, we write a physical operator as a sum
\begin{equation}\label{eq:1}
    \begin{split}
        \mathcal{O}_g\left(x^\mu\right)=\mathcal{O}^{(0)}_\eta\left(x^\mu\right)+\mathcal{O}^{(1)}_\eta\left(x^\mu\right)+\mathcal{O}^{(2)}_\eta\left(x^\mu\right)+\dots
    \end{split}
\end{equation}
where the operators $\mathcal{O}_\eta^{(\alpha)}$ are Lorentz covariant operators such that for any states $\ket{\psi}$, $\ket{\psi'}$ and any point $x^\mu$, we have $\left\vert\bra{\psi'} \mathcal{O}_\eta^{(\alpha)}(x^\mu)\ket{\psi}\right\vert\rightarrow0$ sufficiently quickly as $\alpha\rightarrow\infty$ so that this sum converges.

This construction places restrictions on the physical theory one can write down. In particular, the physical metric $g$ defines the causal structure of the theory and we need to ensure this causal structure is consistent with the standard rules of quantum field theory. For example, we know that any two operators evaluated at spacelike separated points must commute, while operators evaluated at timelike separated points should have non-zero commutator in general (where we evaluate this condition using $\eta$ for the background operators and $g$ for the physical operators). This leads to a consistency condition on the theory. In order for the `Lorentz covariant' approach to hold, we will require that the lightcones defined by $g$ should be contained inside the lightcones defined by  $\eta$. Following \cite{Penrose}, we will denote this condition\footnote{We are using the `mostly minus' signature $(+,-,\dots,-)$} by $g<\eta$, since it tells us that if a curve is timelike with respect to $g$, then it must also be timelike with respect to $\eta$.
\begin{figure}
    \centering
    \includegraphics[scale=0.3]{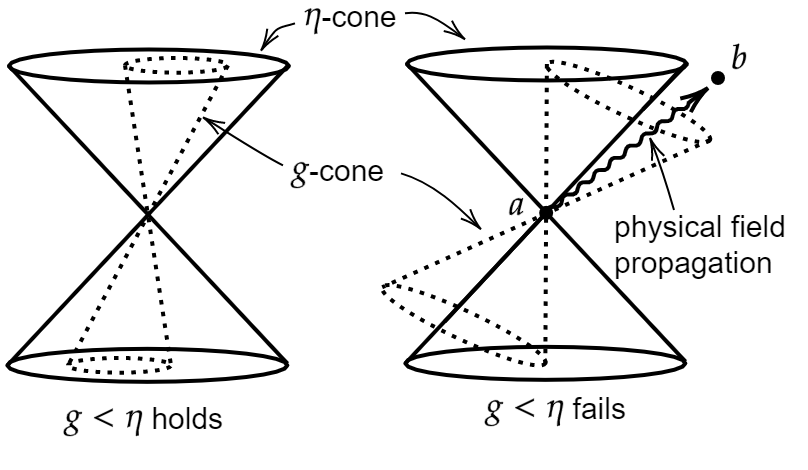}
    \caption{$g<\eta$ is a necessary condition for the `Lorentz covariant' approach to be consistent with QFT}
    \label{fig:g<eta}
\end{figure}
If the condition $g<\eta$ fails, then there are points $a,b$ as shown on the right in Figure \ref{fig:g<eta} which are timelike separated with respect to $g$ but spacelike separated with respect to $\eta$. This means we require the background Lorentz covariant operators defined at these points to commute, but the physical operators should have non-zero commutator in general. This is clearly incompatible with the Lorentz covariant expansion (\ref{eq:1}). 

We now make some definitions which will be used throughout this paper.
\begin{definition}
A \textit{spacetime} $(M,g)$ is a connected manifold $M$ of dimension $d+1$, where $d\geq1$, equipped with a Lorentzian metric $g$ of signature $(1,d)$. We assume also that the metric tensor, $g_{ab}$, is $C^3$ since this will be required for Proposition \ref{thm:Seifert}.
\end{definition}
Following \cite{HawkingEllis}, we make some further definitions
\begin{definition}\label{defn:asymptotically empty and simple}
A time- and space-orientable spacetime $(M,g)$ is 
\textit{asymptotically empty and simple} if there exists a strongly causal spacetime $(\tilde{M},\tilde{g})$ and an embedding map $\lambda:M\rightarrow\tilde{M}$ such that
\begin{enumerate}
\item $\tilde{M}=\lambda(M)\cup\partial\tilde{M}$, where $\partial\tilde{M}$ is the boundary of $\tilde{M}$;
\item $\lambda^*(\tilde{g})=\Omega^2g$, where $\Omega$ is a smooth function on $\tilde{M}$ such that $\Omega>0$ on $\lambda(M)$ and $\Omega=0$, $d\Omega\neq0$ on $\partial\tilde{M}$
\item Every null geodesic in $(M,g)$ is mapped by $\lambda$ to a null geodesic in $(\tilde{M},\tilde{g})$ which has two endpoints on $\partial\tilde{M}$;
\item The Ricci tensor of $g$ vanishes on an open neighbourhood of $\partial\tilde{M}$ in $\tilde{M}$.
\end{enumerate}
\end{definition}
We will mostly omit the embedding map $\lambda$ since the identification between $M$ and $\tilde{M}$ should be clear (the exceptions to this are in Sections \ref{Negative Mass Schwarzschild in Higher Dimensions} and \ref{Kerr Metric}). Definition \ref{defn:asymptotically empty and simple} ensures that a conformal compactification $(\tilde{M},\tilde{g})$ exists for this spacetime. Condition 3 however is extremely restrictive, since it rules out spacetimes containing black hole regions. We will instead consider spacetimes which are \textit{weakly asymptotically empty and simple}.
\begin{definition}\label{defn:weakly asymptotically empty and simple}
A spacetime $(M,g)$ is \textit{weakly asymptotically empty and simple} if there is an asymptotically empty and simple spacetime $(M',g')$ and a neighbourhood $U'$ of $\partial\tilde{M}$ in $M'$ such that $U'\cap M'$ is isometric to an open set $U$ of $M$.
\end{definition}
We will also restrict ourselves to spacetimes containing only one asymptotically flat end. The boundary of the conformally related spacetime $(\tilde{M},\tilde{g})$ can then be split into five parts in the usual way \cite{Wald} - two null surfaces $\mathcal{I}^-$ and $\mathcal{I}^+$ (known as past and future null infinity respectively), spatial infinity denoted $i^0$ as well as past and future timelike infinity denoted $i^-$ and $i^+$ respectively.  All null geodesics which do not enter a black hole region will have past endpoint on $\mathcal{I}^-$ and future endpoint on $\mathcal{I}^+$. Similarly, all timelike geodesics which do not enter a black hole region will have past endpoint on $i^-$ and future endpoint on $i^+$. The definition of a weakly asymptotically empty and simple spacetime includes the Minkowski, Schwarzschild and Kerr spacetimes. 

We will also use units in which $c=G=1$.

In order to understand the condition $g<\eta$ on a spacetime, $(M,g)$, we shall consider the following property: 
\begin{definition}
A spacetime, $(M,g)$, with conformal compactification $(\tilde{M},\tilde{g})$ has the \textit{Penrose property} if any points $p\in\mathcal{I}^-$ and $q\in\mathcal{I}^+$ can be timelike connected (i.e. connected by a smooth timelike curve).
\end{definition}
An equivalent definition given in \cite{Penrose} which makes no mention of conformal compactification is a consequence of the following:
\begin{thm}\label{endless}\textbf{(Penrose \cite{Penrose}, statement that Thm. IV.3 is equivalent to Thm. IV.4)}
A spacetime, $(M,g)$, has the Penrose property if and only if given any two endless timelike curves $\lambda$ and $\mu$ in $M$, there exist points $p\in\lambda$, $q\in\mu$ and a future directed timelike curve from $p$ to $q$.
\end{thm}

\begin{figure} 
    \centering
    \includegraphics[scale=0.3]{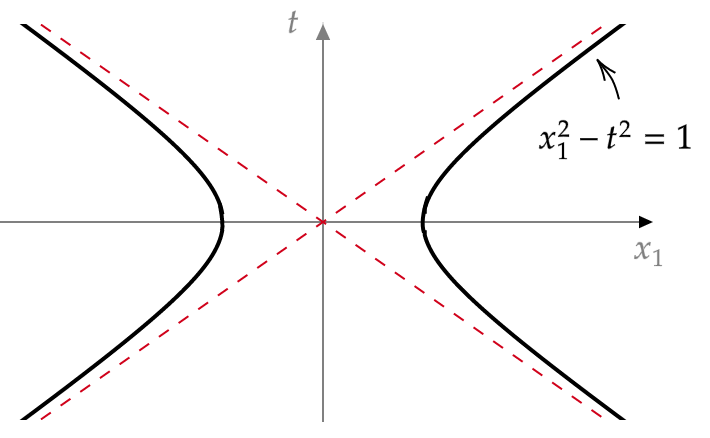}
    
    \caption{Two everywhere spacelike-separated branches of the hyperbola $x_1^2-t^2=1$. These branches tend towards null lines as we approach infinity.}
    \label{fig:hyperbola}
\end{figure}

To begin studying the Penrose property it is helpful to consider the simple example of Minkowski spacetime in $d+1$ dimensions ($d\geq2$):
\begin{thm}\textbf{(Penrose \cite{Penrose}, Thm. IV.5)}\label{Minkowski}
The Penrose property is not satisfied by Minkowski spacetime in $d+1$ dimensions, $\mathbbm{M}^{(1,d)}$, for any $d\geq1$.
\end{thm}
\textbf{Proof:} With respect to co-ordinates $\left(t,x_1,x_2,\dots, x_d\right)$, the Minkowski line element in $d+1$ dimensions can be written as 
\begin{equation}
    ds^2=dt^2-dx_1^2-dx_2^2-\dots-dx^2_d
\end{equation}
Now consider the hyperbola in the $x_1-t$ plane defined by $x_1^2-t^2=1$, $x_2=\dots=x_d=0$. We see (Figure \ref{fig:hyperbola}) that the two branches of this hyperbola are endless timelike curves which are everywhere spacelike separated. We therefore conclude, by Theorem \ref{endless}, that the Penrose property does not hold for Minkowski spacetime in any dimension. \qed

Penrose's motivation for studying the Penrose property in \cite{Penrose} is that it can be related to the consistency condition $g<\eta$ established previously for the `Lorentz covariant' quantum gravity construction:

\begin{prop}\textbf{(Penrose \cite{Penrose}, statement that Thm. IV.5 follows from Thm. IV.3 or Thm. IV.4)}
If the spacetime $(M,g)$ satisfies the Penrose property, then the condition $g<\eta$ fails.
\end{prop}
In the proof of this Proposition, we assume that the compactified manifolds of the physical and background spacetimes are diffeomorphic, at least in some neighbourhood of $\mathcal{I}^-\cup i^0\cup\mathcal{I}^+$ (see definition \ref{defn:weakly asymptotically empty and simple}). We expect this to be the case since $g$ and $\eta$ should have the same asymptotics if the `Lorentz covariant' theory is to be able to describe physical situations such as scattering experiments.

\textbf{Proof: }If $g<\eta$ is satisfied, then curves which are timelike with respect to the physical metric, $g$, must also be timelike with respect to the background metric, $\eta$ (and this is also true for their respective compactifications). So if the Penrose property were to hold for $g$ then it would also hold for $\eta$ -- we can use exactly the same path through the compactified manifolds.  But we have seen that the Penrose property does not hold for Minkowski space which leads to a contradiction and we conclude that $g$ does not satisfy the Penrose property.$\qed$

We will consider the Penrose property as a property in its own right, rather than as a tool to constrain possible theories of quantum gravity. It will turn out that a useful way to prove results relating to this property will be to use a similar trick of comparing the metric of interest to some other metric, defined on the same manifold, whose properties we do understand.

\section{Minkowski Causality} \label{Minkowski Causality}
Before looking at more complicated situations, we should first make sure we fully understand Minkowski spacetime. We have seen from the hyperbola example that in compactified Minkowski space, some points on $\mathcal{I}^-$ cannot be timelike connected to certain points on $\mathcal{I}^+$. To understand exactly which points this applies to, we first note the following lemmas:

\begin{lemma}\label{spherical symmetry}Consider a static, spherically symmetric spacetime, $(M,g)$, which is weakly asymptotically flat at null infinity. Let $\gamma$ be a  timelike curve with endpoints $p\in\mathcal{I}^-$ and $q\in\mathcal{I}^+$. Then we can choose polar co-ordinates $(t,r,\theta_1,...,\theta_{d-2},\phi)$, where $t\in\mathbbm{R}$, $r$ is defined on some subset of $\mathbbm{R}$, $\phi\in[0,2\pi)$ and $\theta_1$,...$\theta_{d-2}\in[0,\pi]$ as well as a timelike curve, $\gamma'$, from $p$ to $q$ such that $\theta_1=...=\theta_{d-2}=\pi/2$ along $\gamma'$.

In other words, it suffices to look for timelike curves between $p$ and $q$ which lie in a submanifold $\Sigma\times S^1\subset M$, where $\Sigma$ 
a 2-dimensional surface with co-ordinates $(t,r)$.
\end{lemma}

\textbf{Proof:} Let the metric take the form
\[
g=A(r)^2dt^2-B(r)^2dr^2-r^2 g_{d-1},
\]
where $g_{d-1}$ is the round metric on a unit $(d-1)$--dimensional sphere $S^{d-1}$ .
We follow \cite{HollandSparling} and consider the conformally re-scaled metric 
\[
\bar{g}=r^{-2}g=g_\Sigma-g_{d-1},
\] 
defined on the product space $\Sigma\times S^{d-1}$, where $g_\Sigma$ is a Lorentzian metric on the surface $\Sigma$. 
An endless non--spacelike curve in the $d+1$ dimensional spacetime therefore corresponds to a pair of curves described by a common parameter, $s$, which we choose to be the arc length of the second curve. One of these is an endless curve in $\Sigma$ and the other is a curve in $S^{d-1}$ (which may be a single point) with endpoints which come from looking at the points on $S^{d-1}$ at which $\gamma$ meets $\mathcal{I}^\pm$ \footnote{Recall that both $\mathcal{I}^+$ and $\mathcal{I}^-$ are topologically $\mathbbm{R}\times S^{d-1}$ - this is a straightforward extension of Prop. 6.9.4 in \cite{HawkingEllis})}. 

If the full curve in $M$ is timelike, then the line element evaluated at each point along it satisfies
$$\bar{ds}^2=ds^2_\Sigma-d\omega^2_{d-1}>0$$
where $ds^2_\Sigma$ and $d\omega^2_{d-1}$ are the line elements along the curves in $(\Sigma,g_\Sigma)$ and $(S^{d-1},g_{d-1})$ respectively.
\begin{figure} 
    \centering
    \includegraphics[scale=0.3]{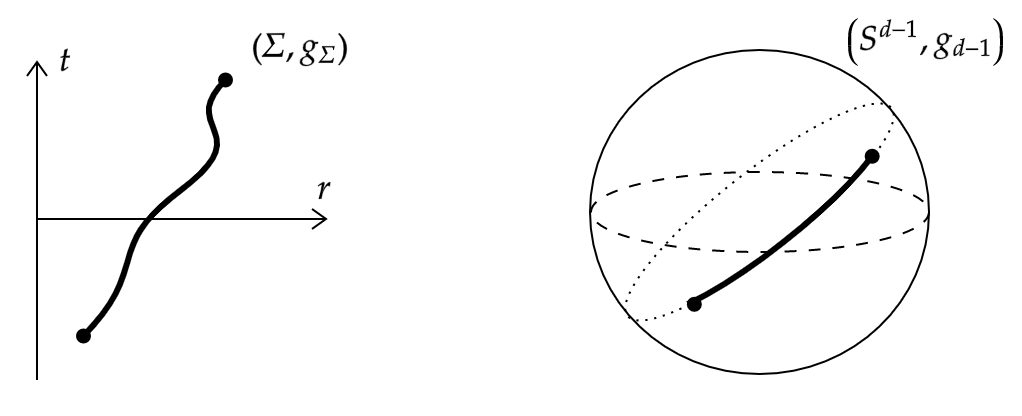}
    \caption{A curve in $d+1$-dimensional, spherically symmetric, static spacetime can be thought of as a pair of curves in $\Sigma$ and $S^{d-1}$.}
    \label{fig:Sparling}
\end{figure}
Now consider the curve in $(S^{d-1},g_{d-1})$. The metric $g_{d-1}$ is a Riemannian metric, so $d\omega^2_{d-1}(s)\geq0$ for each value of the parameter $s$ (with equality corresponding to a radial line segment in $M$). We can modify this curve, keeping the endpoints the same, so that it becomes a geodesic. Since we chose $s$ to be the arc length, this reduces the value of $d\omega^2_{d-1}(s)$ at each value of $s$ and ensures that the curve in the full spacetime remains timelike. Furthermore, the geodesic in $S^{d-1}$ is a segment of a great circle, so we can choose spherical co-ordinates $(\theta_1,\dots,\theta_{d-2},\phi)$ (where $\theta_1,\dots,\theta_{d-2}\in[0,\pi]$, $\phi\in[0,2\pi)$) such that $\theta_1=\dots=\theta_{d-2}=\pi/2$ along the geodesic. This means that the curve in the full spacetime is restricted to the surface $\theta_1=...=\theta_{d-2}=\pi/2$, i.e. a $\Sigma\times S^1$ submanifold of $M$. \qed
\begin{prop}\label{smoothing} \textbf{(Penrose \cite{DiffTop}, Prop. 2.20 and Prop. 2.23)}
Let $\gamma_1$ and $\gamma_2$ be future pointing causal (i.e. timelike or null) geodesics from $a$ to $b$ and from $b$ to $c$ respectively. Then either there exists a smooth timelike curve from $a$ to $c$ or the union of these geodesics, $\gamma=\gamma_1\cup\gamma_2$, is itself a null geodesic. 
\end{prop}
\begin{figure} 
  \centering
        \includegraphics[scale=0.3]{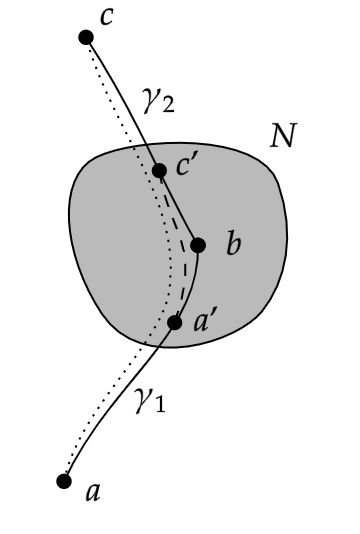}
    \caption{Assume $\gamma_1\cup\gamma_2$ is not a single null geodesic. Then we choose the points $a'$ and $c'$ to lie in a normal neighbourhood of $b$, so there exists a timelike geodesic between them (dashed line). Thus we have connected $a$ to $c$ using three causal geodesic segments, where in particular one of these segments is timelike. This can be smoothed to give a timelike curve from $a$ to $c$ (dotted line).}
    \medskip
    \medskip\medskip
    \label{fig:smoothing}
\end{figure}
We now return to considering Minkowski spacetime in $d+1$ dimensions, $\mathbbm{M}^{(1,d)}$. Since this spacetime is spherically symmetric, Proposition \ref{smoothing} ensures that if two points can be connected by a timelike curve then we can choose this curve to lie in a 3-dimensional surface with induced metric equal to that of $\mathbbm{M}^{(1,2)}$. It therefore suffices to consider only the compactification of Minkowski spacetime in 2+1 dimensions. The Minkowski line element in $2+1$ dimensions in polar co-ordinates is
\begin{equation}\label{eqn:Mink}
    ds^2=dt^2-dr^2-r^2d\phi^2
\end{equation}
We define retarded and advanced time co-ordinates respectively by 
\begin{equation}\label{eqn:MinkRet}
    \begin{split}
        u&=t-r, \quad
        v=t+r
    \end{split}
\end{equation}
We then compactify the metric by defining
\begin{equation}\label{eqn:retardedadvanced}
u=\tan P, \quad
v=\tan Q
\end{equation}
Finally, we define new co-ordinates 
\begin{equation}
 T=Q+P\in(-\pi,\pi), \quad
        \chi=Q-P\in[0,\pi)
    \end{equation}
which we can think of as ``time" and ``radial" co-ordinates respectively in the compactified spacetime.

We then consider the conformally related metric $\tilde{g}$, given by
$$\tilde{g}=\Omega^2g=\left(2\cos P\cos Q\right)^2g$$
This gives the line element
\begin{equation}\label{eqn:compactMink}
    \Tilde{ds}^2=dT^2-d\chi^2-\sin^2\chi d\phi^2
\end{equation}

\begin{figure} 
 \minipage{0.48\textwidth}
    \centering
    \includegraphics[scale=0.2]{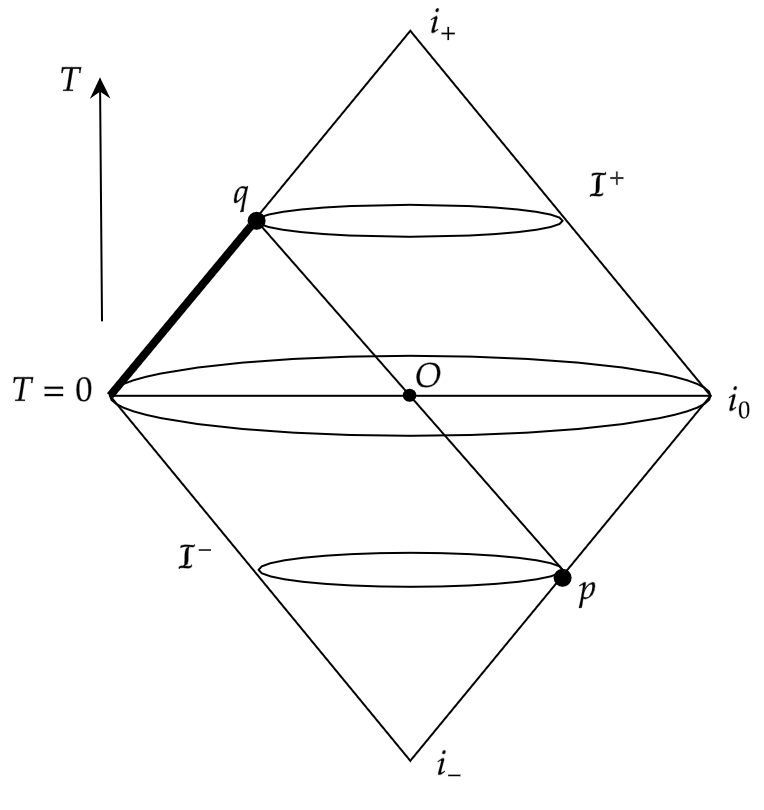}
    \caption{The points $p\in\mathcal{I}^-$ and $q\in\mathcal{I}^+$ are connected by a radial null geodesic through the origin. The point $p$ can be timelike connected to all points in $\mathcal{I}^+$ except those that lie to the past of $q$ and on the same null generator of $\mathcal{I}^+$. These points correspond to the shaded interval.}
    \label{fig:doublecone}
\endminipage
 \minipage{0.04\textwidth}
    \includegraphics[scale=0.05]{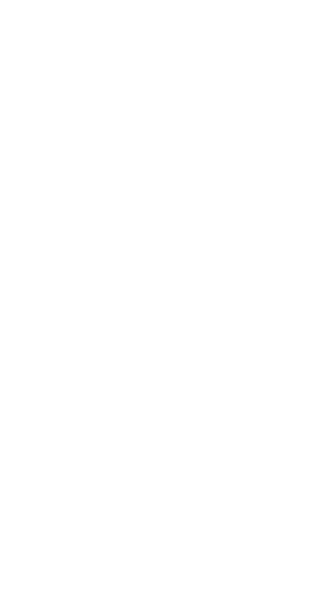}
 \endminipage
  \minipage{0.48\textwidth}
    \centering
    \includegraphics[scale=0.2] {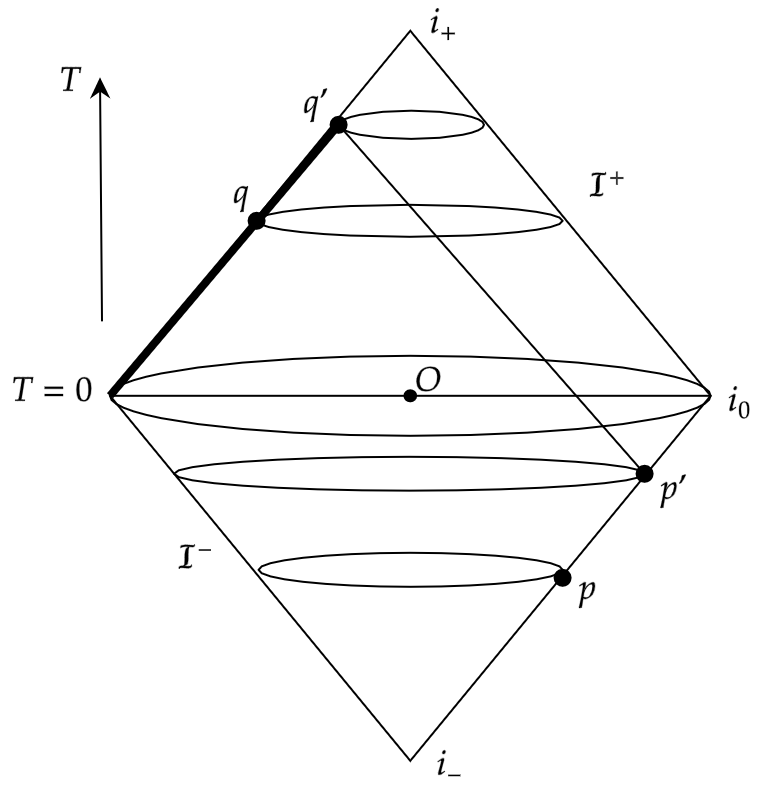}
    \caption{The point $p'\in\mathcal{I}^-$ can be connected to all points on $\mathcal{I}^+$ except those that lie to the past of $q'$ and on the same null generator of $\mathcal{I}^+$ (i.e. the shaded interval), where $q'$ is the unique future endpoint of null geodesics from $p'$ which enter $M$.}
    \label{fig:shift}
    \endminipage
\end{figure}

In 2+1 dimensions, the compactified manifold, $\tilde{M}$, is a double cone, as shown in Figure \ref{fig:doublecone}, where we can think of time as moving upwards. 
The antipodal points $p$ and $q$ in Figure \ref{fig:doublecone} correspond to the past and future endpoints of different branches of the hyperbola shown in Figure \ref{fig:hyperbola}. They have $T$ and $\chi$ co-ordinates $(T,\chi)=(-\frac{\pi}{2},\frac{\pi}{2})$, $(\frac{\pi}{2},\frac{\pi}{2})$ respectively. The radial null geodesic between these points is a straight line in this diagram. Moreover, any null geodesic from $p$ which enters the interior spacetime, $M$, has future endpoint at $q$. For a general point $p'\in\mathcal{I}^-$, there is a unique $q'\in\mathcal{I}^+$ which is the future endpoint of null geodesics from $p'$ which enter $M$. The points $p'$ and $q'$ can be obtained from $p$ and $q$ by a simple translation in $T$ and $\phi$.

\begin{prop}\label{prop:Mink}
Let $p\in\mathcal{I}^-$ and let $q$ be the unique point in $\mathcal{I}^+$ which is connected to $p$ by a null geodesic which enters $M$. Then the only points in $\mathcal{I}^+$ which cannot be timelike connected to $p$ are the points in $\mathcal{I}^+$ which can be reached from $q$ by a past pointing null geodesic, including the point $q$ itself. That is, for any $p\in\mathcal{I}^-$
\begin{equation}
    I^+(p)\cap\mathcal{I}^+=\mathcal{I}^+\setminus \left(J^-(q)\cap\mathcal{I}^+\right)
\end{equation}
where $q\in\mathcal{I}^+\cap\partial J^+(p)$ can be connected to $p$ by a null geodesic which does not pass through $i^0$ (and hence enters $M$) and where we define the following sets
\begin{equation}
    \begin{split}
        J^+(p)&=\{q\in\tilde{M}|\exists\text{ a smooth future-directed causal curve from $p$ to $q$}\}\\
        J^-(p)&=\{q\in\tilde{M}|p\in J^+(q)\}\\
        I^+(p)&=\{q\in\tilde{M}|\exists\text{ a smooth future-directed timelike curve from $p$ to $q$}\}\\
        I^-(p)&=\{q\in\tilde{M}|p\in I^+(q)\}\\
    \end{split}
\end{equation}
For non-compact spacetimes, $(M,g)$, we can define the same sets as subsets of $M$ rather than as subsets of $\tilde{M}$.
\end{prop}
This proposition is illustrated in Figure \ref{fig:shift}. In particular, we note that antipodal points close to spacelike infinity cannot be timelike connected.

\textbf{Proof:} Let $p\in\mathcal{I}^-$ have co-ordinates $(T,\chi,\phi)=(-\frac{\pi}{2},\frac{\pi}{2},0)$, so $q$ has co-ordinates $(\frac{\pi}{2},\frac{\pi}{2},\pi)$. This is shown in Figure \ref{fig:doublecone}. These points correspond to the past and future endpoints of separate branches of the hyperbola shown in Figure \ref{fig:hyperbola}. A general point $p'\in\mathcal{I}^-$ is related to $p$ by a translation in $T$ and $\phi$. Since the compactified metric (\ref{eqn:compactMink}) does not depend on these co-ordinates, it suffices to prove the Proposition for this point $p$. The same argument would then apply to general $p'\in\mathcal{I}^-$, only with the point $q\in\mathcal{I}^+$ shifted in the same way to become $q'$. This is shown in Figure \ref{fig:shift} for a translation in $T$ (we assume without loss of generality that $\phi=0$ at $p'$). 

\begin{figure} 
  \minipage{0.48\textwidth}
        \includegraphics[scale=0.2]{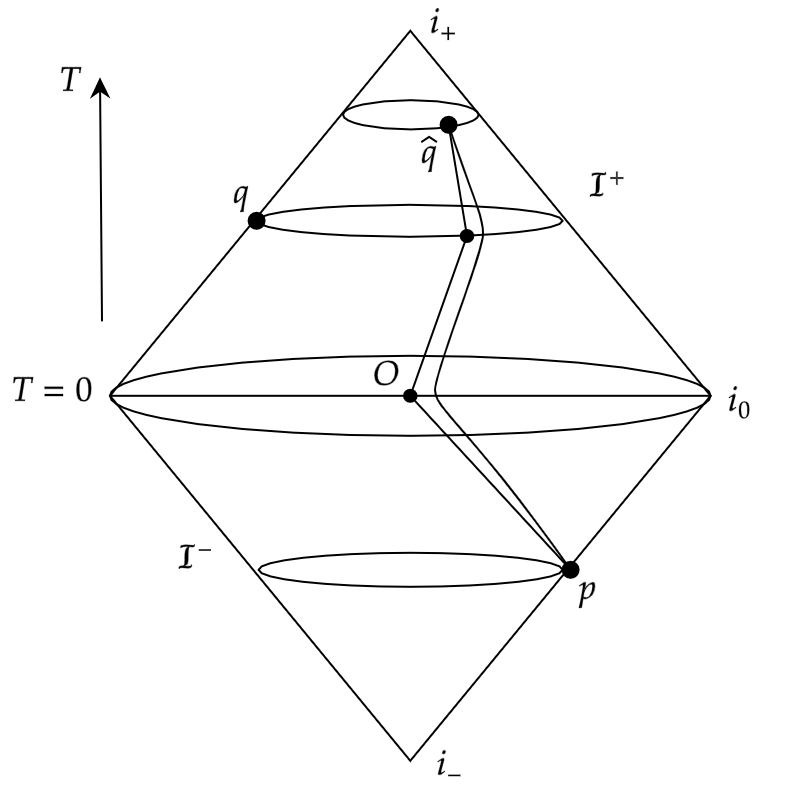}
    \caption{$p$ can be timelike connected to any point on $\mathcal{I}^+$ with $T$ co-ordinate greater than or equal to that of $q$ (other than $q$ itself). The path shown consists of two radial null lines through the origin as well as a null line up $\mathcal{I}^+$. Proposition \ref{smoothing} then implies the existence of a smooth timelike curve between $p$ and $\hat{q}$.}
    \label{fig:doublecone2}
    \endminipage
     \minipage{0.04\textwidth}
      \includegraphics[scale=0.05]{blank.png}
      \endminipage
    \minipage{0.48\textwidth}
 \centering
    \includegraphics[scale=0.195]{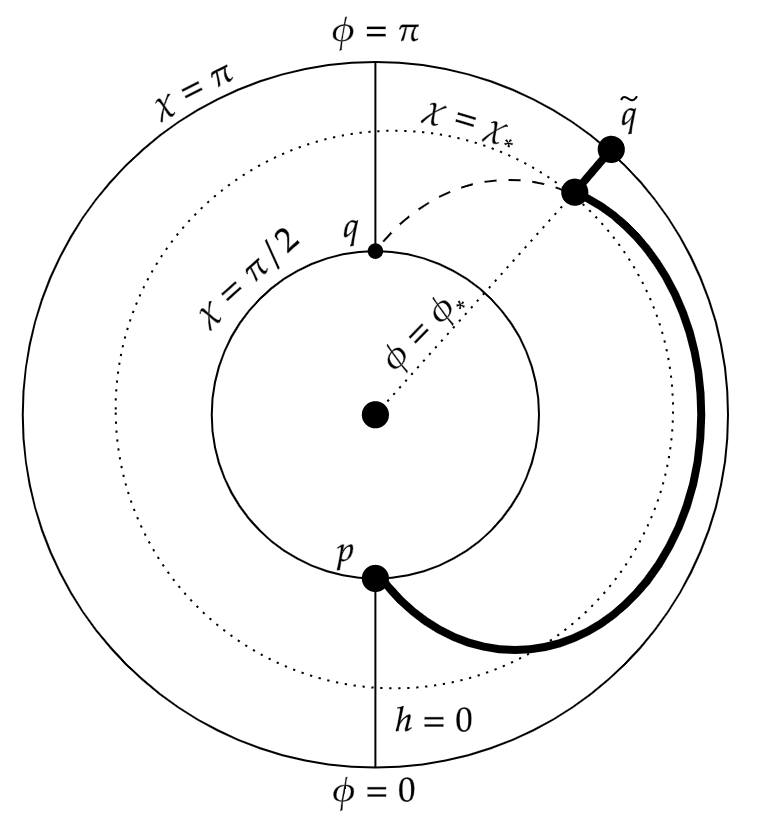}
    \caption{Double cone of compactified Minkowski spacetime viewed from above.}
    \medskip
    \medskip
    \medskip
    \medskip
    \medskip
    \medskip
    \medskip
    \medskip
    \medskip
    \label{fig:conefromabove}
    \endminipage
\end{figure}

First we note that it is clear none of the shaded points in Figure \ref{fig:doublecone} lie in the interior of the future lightcone from $p$. Next observe that it is straightforward to find a timelike curve from $p$ to points such as $\hat{q}$ in Figure \ref{fig:doublecone2} which lie on $\mathcal{I}^+$ with $T$-value greater than or equal to that of $q$ (and excluding $q$ itself). We can use two radial null geodesics intersecting at the origin $O$ as well as a radial null geodesic up $\mathcal{I}^+$ if necessary. Since $\hat{q}\neq q$, the total curve is not a null geodesic, so we can use Proposition \ref{smoothing} to deduce that there exists a smooth timelike curve from $p$ to $\hat{q}$. By using a curve of constant $\phi$ which is also a straight line in $(T,\chi)$ co-ordinates, we are able to timelike connect $p$ to any point in $\mathcal{I}^+$ which has the same $\phi$ co-ordinate as $p$. This means we can concentrate on points on $\mathcal{I}^+$ with larger $\chi$ co-ordinates than $p$, i.e. which are closer to spacelike infinity, and whose $\phi$ co-ordinate differs from that of $p$ by a value in $(0,\pi)$ (without loss of generality). 

It is helpful to look at the double cone from above, as in Figure \ref{fig:conefromabove}. The circles shown correspond to surfaces $\chi=$constant, with the larger circle $\chi=\pi$ representing spatial infinity. 
To show that all non-antipodal points can be timelike connected, we will follow an argument similar to the one used by Penrose in $\cite{Penrose}$. This argument, which will be used again in section \ref{Positive Mass Schwarzschild in 3+1 Dimensions}, relies on repeated applications of Proposition \ref{smoothing} to a curve which is composed of segments of null geodesics. More specifically, the curve is based on several radial null geodesics, as well as a single non-radial one, which we use to achieve the desired change in the angular co-ordinate $\phi$.

From the null geodesic equations, we have
\begin{equation}
\begin{split}
\dot T &= 1 \text{ (wlog)}\\
\sin^2(\chi)\dot\phi&=h=\text{constant}\\
 \dot\chi^2&=1-\frac{h^2}{\sin^2\chi}
    \end{split}
\end{equation}
where overdots denote differentiation with respect to some parameter which we have normalised so that $\dot{T}=1$.

We can integrate these equations to find explicit expressions for $\chi$ and $\phi$ involving $T$. Firstly we have
\begin{equation}
   \tan^2T=\frac{\sin^2\chi-h^2}{\cos^2\chi}
\end{equation}
where we have chosen the geodesic to intersect $\mathcal{I}^-$ at $(T,\chi)=(-\pi/2,\pi/2)$. This ensures that the maximum value of $\chi$ is achieved at $T=0$ and that the curve intersects $(T,\chi)=(\pi/2,\pi/2)$. Using this, we can solve for $\phi$ along the curve (assuming $h\neq0$). We have
\begin{equation}\label{eqn:phi}
    \begin{split}
        \sin^2\chi\dot{\phi}&=h\\
        \implies\dot{\phi}&=\frac{h\left(1+\tan^2T\right)}{h^2+\tan^2T}\\
        \implies\phi&=\tan^{-1}\left(\frac{\tan T}{h}\right)+\frac{\pi}{2}\\
    \end{split}
\end{equation}
Where we have chosen our geodesic to have $\phi=0$ at $\mathcal{I}^-$. This ensures that $\phi=\pi$ on $\mathcal{I}^+$ for $h\neq0$.

Consider a point $\tilde{q}\in\mathcal{I}^+$ at which $\chi=\chi_*\in(\pi/2,\pi)$ and $\phi=\phi_*\in(0,\pi)$. It suffices to show that a null geodesic with past endpoint at $p$ reaches a point with these same values of the $\chi$ and $\phi$ co-ordinates, since we could then reach $\tilde{q}$ by following this null geodesic until we reach such a point before switching to a timelike geodesic of constant $\chi, \phi$. This construction is shown as the bold path in Figure \ref{fig:conefromabove}. Finally we could then apply Proposition \ref{smoothing} to obtain a smooth timelike curve with the same endpoints at $p$ and $\tilde{q}$. 

Re-arranging equation (\ref{eqn:phi}), the null geodesic intersects the required point provided we choose
$$h^2=\frac{\sin^2\chi_*}{1+\cos^2\chi_*\tan^2(\phi_*-\pi/2)}$$
where we choose the positive sign for $h$ since we assumed $\phi_*\in(0,\pi)$.

Choosing this value of $h$ gives us the required null geodesic and we conclude that points in $\mathcal{I}^-$ with $\chi=T=\frac{\pi}{2}$ are timelike connected to all non-antipodal points in $\mathcal{I}^+$.$\qed$
\section{The Penrose Property in Schwarzschild Spacetime} 
We now consider the effect of adding mass. In particular, we will consider the Schwarzschild spacetime in various dimensions and for both positive and negative mass. 

\subsection{Schwarzschild Spacetime in 2+1 Dimensions}\label{Schwarzschild of Dimension 2+1}
\subsubsection{Positive Mass}We begin by focusing on the 2+1 dimensional case, where we prove the following result:
\begin{prop}
\label{prop41}
The Penrose property is satisfied by the Schwarzschild spacetime of mass $m$ in 2+1 dimensions if and only if $m>0$. 
\end{prop}
\textbf{Proof:} We first note that the failure of the Penrose property when $m=0$ has already been established in section \ref{Minkowski Causality} (since the spacetime in this case is Minkowski with a point removed). We now consider the case where the mass is non-zero.

In 2+1 dimensions, the symmetries of the Riemann tensor are highly constraining, and we have \cite{Star}
\begin{equation}
    R_{abcd}=\left(R^{mn}-\frac{1}{2}Rg^{mn}\right)\epsilon_{mab}\epsilon_{ncd}
\end{equation}
where $\epsilon_{abc}$ is the fully anti-symmetric tensor of rank 3 with $\epsilon_{123}=1$.

Away from sources, the Einstein equations imply $R_{ab}=0$ and hence spacetime is flat. This means that gravity in 2+1 dimensions is different to the theory in higher dimensions. In particular, the curvature is fully determined by the Einstein equations and there is no propagating field of gravity.

The stress-energy tensor for a point source of mass $m$ at the origin is
$$T^{00}=m\delta(\boldsymbol{r}),\text{ } T^{0i}=T^{ij}=0$$
The spacetime is static and spherically symmetric, so we can write the metric in isotropic co-ordinates as
$$ds^2=N(\textbf{r})dt^2-\varphi(\textbf{r})(dr^2+r^2d\phi^2)$$
Solving the Einstein equations in vacuum $\left(R_{ab}=0\right)$, we find (\cite{DJtH}, Eq. 2.4) that $N$ is a constant, which we can choose to be 1 by a re-scaling of $t$. Next we solve for $\varphi(\textbf{r})$ to find (\cite{DJtH}, Eq. 2.6)
$$\varphi(\textbf{r})=r^{-8m}$$
Making the change of variables\footnote{We are actually restricting here to $m<\frac{1}{4}$. The case $m>\frac{1}{4}$ corresponds to a `mass at infinity', while $m=\frac{1}{4}$ gives a cylinder in the spatial dimensions. See \cite{DJtH} for a discussion of this. We will not consider these cases further.}  $\tilde{r}=\alpha^{-1}r^\alpha$, $\tilde{\phi}=\alpha\phi$, where $\alpha=1-4m$, we arrive at the metric
\begin{equation}
    ds^2=dt^2-d\tilde{r}^2-\tilde{r}^2
d\tilde{\phi}^2
\end{equation}
We recognise this as the Minkowski metric (\ref{eqn:Mink}). The important thing to note is that the co-ordinates $(t,r,\phi)$ do not cover all of $\mathbbm{R}^3$. This is because the range of $\tilde{\phi}$ is now $0\leq\tilde{\phi}<2\pi\alpha$, so there is a conical singularity at the origin.

If $m>0$, the range of $\tilde{\phi}$ is strictly less than $2\pi$. We have therefore introduced a deficit angle 
$$\tilde{\phi}_{deficit}=8\pi m$$
This means that any two points on $\mathcal{I}^-$ and $\mathcal{I}^+$ will have $\tilde{\phi}$ co-ordinates which differ by strictly less than $\pi$ (mod $2\pi$), so we have effectively removed all the antipodal points. By the arguments presented in section \ref{Minkowski Causality} (in particular Proposition \ref{prop:Mink} and its extension to general points on $\mathcal{I}^-$), this means we are able to connect any pair of points on $\mathcal{I}^-$ and $\mathcal{I}^+$ with a timelike curve, so the Penrose property is satisfied.
\begin{figure} 
\minipage{0.48\textwidth}
    \centering
    \includegraphics[scale=0.45] {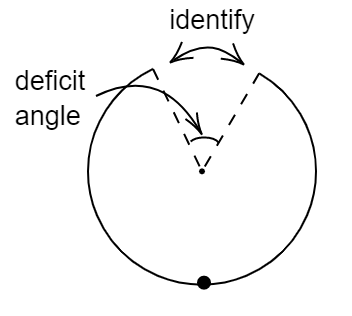}
    \label{deficit}
    \caption{The deficit angle resulting from inserting a positive point mass removes antipodal points.}
    \endminipage
 \minipage{0.04\textwidth}
 \includegraphics[scale=0.05]{blank.png}
  \endminipage
\minipage{0.48\textwidth}
    \centering
    \includegraphics[scale=0.2] {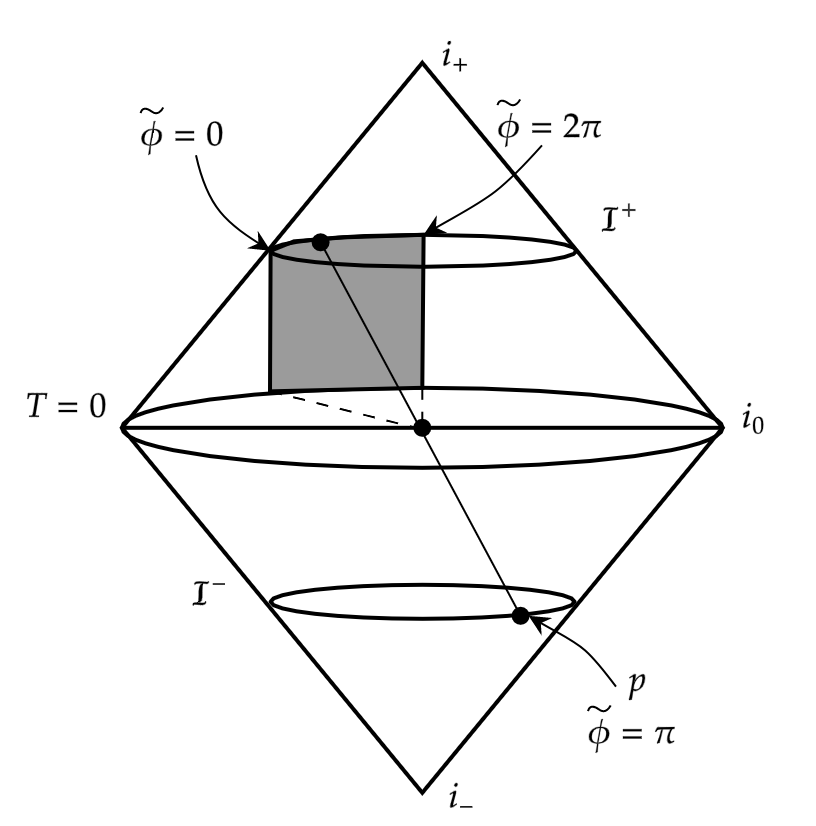}
    \caption{The surplus angle resulting from inserting a negative point mass means that given any point on $\mathcal{I}^-$ there is now a wedge of points on $\mathcal{I}^+$ which cannot be reached by a timelike curve.}
    \label{fig:surplus}
    \endminipage
\end{figure}

On the other hand, inserting a negative mass yields the opposite effect. We now have a surplus angle, so $\tilde{\phi}$ has a range strictly greater than $2\pi$. The result of this is that given any initial point $p\in\mathcal{I}^-$, there is a wedge of points on $\mathcal{I}^+$ which cannot be reached by a timelike curve (see Figure \ref{fig:surplus}). At these points, the $\tilde{\phi}$ co-ordinate differs by strictly greater than $\pi$ from the value at $p$. The Penrose property is not satisfied by this negative mass spacetime. In fact, the situation is worse than in Minkowski, since the set of points on $\mathcal{I}^+$ which cannot be reached by a timelike curve from $p$ now forms a 2-dimensional area (as opposed to the 1 dimensional line segment we found in Minkowski spacetime).$\qed$

So it appears that the Penrose property is linked to the presence of positive 
mass\footnote{This link, and some of its consequences has been  explored in \cite{Chr, DT19, Ga, GW, PSW}. }. This is not wholly unsurprising since we expect a positive mass source to focus and retard nearby geodesics (and to correspondingly time-advance geodesics passing far from the source as described in \cite{PSW}). This should make it easier for timelike curves to reach antipodal points near $i^0$. The results in 2+1 dimensions, along with this intuition, suggest that we should look at positive mass spacetimes if we want to satisfy the Penrose property. 

\subsection{Schwarzschild Spacetime in 3+1 Dimensions}\label{Schwarzschild of Dimension 3+1}
We now consider the Schwarzschild solution in 3+1 dimensions. As outlined in Theorem A, we will find that the Penrose property is satisfied when the mass is $m>0$ but not when $m<0$ (the case $m=0$ corresponds to Minkowski spacetime and has already been considered in section \ref{Minkowski Causality}).
\subsubsection{Positive Mass}
\label{Positive Mass Schwarzschild in 3+1 Dimensions}
 The positive mass Schwarzschild metric in 3+1 dimensions is
\begin{equation}\label{eqn:Schwarzschild}
    \begin{split}
        ds^2=V(r)dt^2-\frac{dr^2}{V(r)}-r^2\left(d\theta^2+\sin^2\theta d\phi^2\right)
    \end{split}
\end{equation}
where $V(r)=1-\frac{r_s}{r}$ and $r_s=2m>0$ ($m$ is the mass of the spacetime).

We can conformally compactify this metric using the same procedure as was used in section \ref{Minkowski Causality} for Minkowski spacetime, the only difference being we now use the tortoise co-ordinate, $r_*$, which satisfies
\begin{equation}\label{eqn:tortoise}
\begin{split}
     dr_*&=\frac{dr}{V(r)}\\
\end{split}
\end{equation}
in order to define the retarded and advanced time co-ordinates
\begin{equation}\label{eqn:SchwRet}
    \begin{split}
        u&=t-r_*=t-r-r_s\log(r/r_s-1)\\
v&=t+r_*=t+r+r_s\log(r/r_s-1)
    \end{split}
\end{equation}
where we have set the arbitrary integration constant in the definition of $r_*$ (equation (\ref{eqn:tortoise})) to zero.

Just as in section \ref{Minkowski Causality}, we define compactified ``time'' and ``radial'' co-ordinates $T\in(-\pi,\pi)$ and $\chi\in[0,\pi)$ to be
\begin{equation}\label{eqn:Tchi}
    T=\arctan v+\arctan u, \quad \chi=\arctan v-\arctan u
\end{equation}
The conformally compactified metric is then
\begin{equation}
\label{eqn:compactified}
    \begin{split}
        \Tilde{ds}^2=\Omega^2ds^2=dT^2-d\chi^2-\frac{r^2}{V(r)r_*^2}\sin^2\chi\left(d\theta^2+\sin^2\theta d\phi^2\right),
    \end{split}
\end{equation}
where $\Omega=\frac{2\cos \left(\frac{T+\chi}{2}\right)\cos \left(\frac{T-\chi}{2}\right)} {\sqrt{V(r)}}$.

The compactified manifold for the region of Schwarzschild spacetime $\{r\geq r_{\#}\}$, where $r_\#$ is the unique solution to $r_\#+r_s\log(r_{\#}/r_s-1)=0$, is the same as the compactified Minkowski manifold, namely the (higher dimensional) cone $$\tilde{M}=\{(T,\chi)\in[-\pi,\pi]\times[0,\pi]:|T|\leq\pi-\chi\}\times S^2$$
Note that $r_\#>r_s$, so $V(r)>0$ at all points in the compactified manifold.
\begin{thm}\textbf{(Penrose \cite{Penrose})}\label{thm:51}
Positive mass Schwarzschild spacetime in $3+1$ dimensions satisfies the Penrose property.
\end{thm}
\textbf{Proof:} We follow the method used by Penrose in \cite{Penrose} to connect points $p\in\mathcal{I}^-$ and $q\in\mathcal{I}^+$. We also fill in some details which are ommitted from \cite{Penrose}, in particular we show how the positivity of mass ensures that we can follow null geodesics in order to reach antipodal points (see equation (\ref{eqn:deltaphi})). 

Just as we have done in 2+1 dimensions, this method uses a piecewise null geodesic construction, with the existence of a smooth timelike curve from $p$ to $q$ guaranteed by Proposition \ref{smoothing}. We choose co-ordinates $(t,r,\theta,\phi)$ such that $\theta=\pi/2$ at $p$ and $q$. The Schwarzschild metric is static and spherically symmetric, so by Lemma \ref{spherical symmetry} it suffices to consider curves restricted to the hypersurface $\theta=\pi/2$. All curves referred to in this section will be assumed to lie in this plane. Up to shifts in $t$ and $\phi$, Schwarzschild geodesics in this plane which are null and inextendible (so in particular $r\geq r_s$) are uniquely determined by their impact parameter, $R$, defined to be the minimal value of $r$ along the curve (which we assume is obtained at $t=0$). The geodesic Lagrangian
is
\[
L=V(r)\dot{t}^2-V(r)^{-1}\dot{r}^2-r^2\dot{\phi}^2
\]
so that, in the null case,
\[
L=0, \quad h=-\frac{1}{2}\frac{\partial L}{\partial \dot{\phi}}=r^2\dot{\phi}=\mbox{const}, \quad
\frac{1}{2}\frac{\partial L}{\partial \dot{t}}=V(r)\dot{t}=\mbox{const}=1,
\]
where the last equation is our choice of the parametrisation.

For such a geodesic, which in particular must have non-zero angular momentum, $h$, we have:
\begin{equation}
    \begin{split}
        \dot{r}^2&=1-\left(1-\frac{r_s}{r}\right)\frac{h^2}{r^2} \\
        &\leq1-\left(1-\frac{r_s}{R}\right)\frac{h^2}{r^2}\\
        &=1-\frac{R^2}{r^2}
    \end{split}
\end{equation}
where we use the fact that $\dot{r}=0$ at $r=R$ to solve for $h^2=R^2\left(1-\frac{r_s}{R}\right)^{-1}$.

\begin{figure} 
  \minipage{0.38\textwidth}
  \centering
      \includegraphics[scale=0.23]{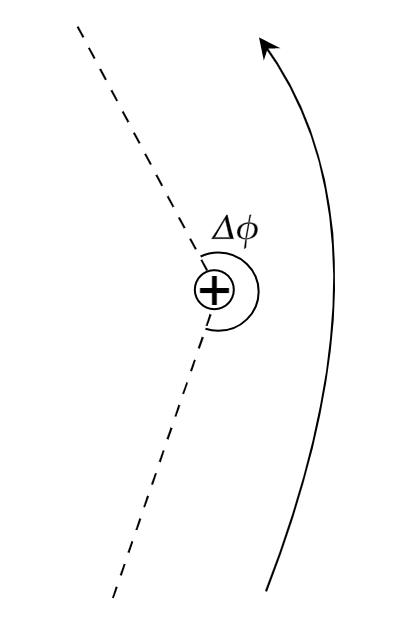}
    \caption{$|\Delta\phi|>\pi$ along null geodesics in positive mass Schwarzschild.
    \medskip
    \medskip
    \medskip
    \medskip}
    \label{fig:positivemass}
    \endminipage
     \minipage{0.04\textwidth}
      \includegraphics[scale=0.05]{blank.png}
      \endminipage
    \minipage{0.58\textwidth}
 \centering
   \includegraphics[scale=0.3]{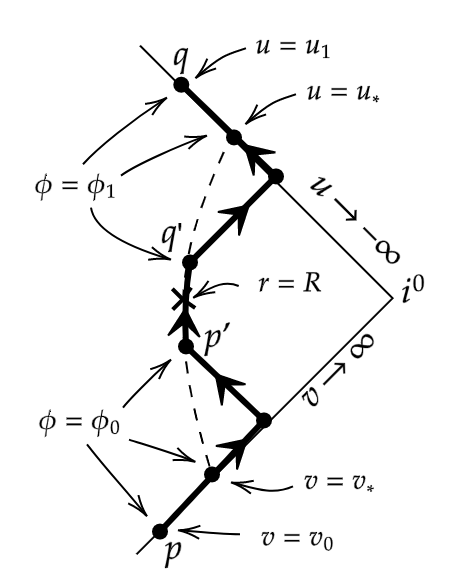}
    \caption{Construction used in \cite{Penrose} to timelike connect points $p\in\mathcal{I}^-$ and $q\in\mathcal{I}^+$. This is based on using a null geodesic (dashed line) to vary the angular co-ordinate $\phi$ (working in the hypersurface $\theta=\frac{\pi}{2}$). We then complete the path using null geodesics with zero angular momentum. The existence of a smooth timelike curve from $p$ to $q$ is then guaranteed by Proposition \ref{smoothing}.}
    \label{fig:Penrose}
    \endminipage
\end{figure}

Using this, we can obtain a lower bound for $|\Delta\phi|$, the magnitude of the change in $\phi$ along this null geodesic:
\begin{equation}\label{eqn:deltaphi}
\begin{split}
        \dot{\phi}&=\frac{h}{r^2}\\
        \implies|\Delta\phi|&=2R\left(1-\frac{r_s}{R}\right)^{-1/2}\int_R^\infty\frac{dr}{\dot{r}r^2}\\
        &\geq2R\left(1-\frac{r_s}{R}\right)^{-1/2}\int_R^\infty\frac{dr}{\left(1-R^2/r^2\right)^{1/2}r^2}\\
        &=\left[-2\left(1-\frac{r_s}{R}\right)^{-1/2}\tan^{-1}\left(\frac{R}{\sqrt{r^2-R^2}}\right)\right]_{r=R}^\infty\\
        &=\left(1-r_s/R\right)^{-1/2}\pi\\
        &\geq\pi
        \end{split}
\end{equation}
Therefore $|\Delta\phi|\geq\pi$ along a null geodesic in the $\theta=\frac{\pi}{2}$ plane. This was to be expected near a positive mass source  (see Figure \ref{fig:positivemass}).

Using this fact, we can connect any points on $\mathcal{I}^-$ and $\mathcal{I}^+$ which lie outside the intervals between the ends of this null geodesic and spatial infinity. The construction is illustrated in Figure \ref{fig:Penrose}. In this figure we are interested in finding a timelike curve between the point $p\in\mathcal{I}^-$ (with $v=v_0$, $\phi=\phi_0$) and the point $q\in\mathcal{I}^+$ (with $u=u_1$, $\phi=\phi_1$). The details of this construction are as follows.

Suppose we are able to choose the impact parameter $R$ such that the past and future endpoints of the null geodesic 
have $v=v_*>v_0$ and $u=u_*<u_1$ respectively. Along this geodesic, we choose points $p'$ and $q'$ (with $q'\in J^+(p')$) at which $\phi=\phi_0$ and $\phi=\phi_1$ respectively (we know this is possible because we have shown that $|\Delta\phi|\geq\pi$ along this null geodesic and we can choose the angular momentum $h$ to be positive or negative). We use this segment of geodesic to vary the value of $\phi$. We connect $p$ to $p'$ using two null geodesics with zero angular momentum (so $\phi$ is constant along them), one of which runs along null infinity towards $i^0$. We connect $q$ to $q'$ similarly. This gives us a piecewise null geodesic curve from $p$ to $q$. We deduce, using repeated applications of Proposition \ref{smoothing}, that a smooth timelike curve from $p$ to $q$ must exist.
So we have reduced the Penrose property to a problem concerning the end points of null geodesics in Schwarzschild. The construction described above will allow us to connect all pairs of points on $\mathcal{I}^-$ and $\mathcal{I}^+$ if we are able to choose null geodesic endpoints arbitrarily close to spatial infinity. This also demonstrates how the Penrose property is a property of neighbourhoods of spatial infinity. If we know that we can connect all points arbitrarily close to $i^0$ then this construction shows that the Penrose property is satisfied.

Points on $\mathcal{I}^-$ and $\mathcal{I}^+$ can be labelled by co-ordinates $(v,\theta,\phi)$ and $(u,\theta,\phi)$ respectively. The $u$ co-ordinate, $u_*$, at the future endpoint of a non-radial null geodesic, chosen to be symmetric about $t=0$, is given in terms of the $u$ co-ordinate at $r=R$, denoted $u_R$, by
\begin{equation}
    \begin{split}
      u_*&=u_R+\int_R^\infty\frac{du}{dr}dr\\
      &=-R-r_s\log\left(\frac{R}{r_s}-1\right)+\int_R^\infty\frac{du}{dr}dr
    \end{split}
\end{equation}
In \cite{Penrose}, Penrose shows that
$$\int_R^\infty\frac{du}{dr}dr=R+\mathcal{O}(1)$$
where $\mathcal{O}(1)$ denotes terms which are bounded as $R\rightarrow\infty$. This proves that $u_*\rightarrow-\infty$ as $R\rightarrow\infty$. A similar argument also shows that $v_*\rightarrow\infty$ as $R\rightarrow\infty$. This is exactly what we wanted. We can therefore use this construction to connect any pairs of points on $\mathcal{I}^-$ and $\mathcal{I}^+$ using a timelike curve. We simply need to use a geodesic with sufficiently large impact parameter. So we see that the Penrose property holds, and as a result the perturbative quantum gravity consistency condition $g<\eta$ described in section \ref{`Lorentz Covariant' Quantum Gravity}, is not satisfied.$\qed$
\subsubsection{Negative Mass}
\label{Negative Mass Schwarzschild in 3+1 Dimensions}
We now consider the Schwarzschild metric in 3+1 dimensions with negative mass. The compactified metric is now as in (\ref{eqn:compactified}) except with\begin{equation}\label{eqn:negativemasstortoise}
    \begin{split}
        V(r)&=1+\frac{r_s}{r}\\
        \implies r_*&=r-r_s\log(r/r_s+1)
    \end{split}
\end{equation}
and the compactified manifold covers the whole spacetime $\{r>0\}$ outside the naked singularity. This naked singularity will not be relevant for our discussion because, as explained in the positive mass case, we are concerned only with properties of the spacetime near $i^0$ (and hence at large values of $r$).
\begin{prop}
\label{prop61}
Negative mass Schwarzschild spacetime in $3+1$ dimensions does not satisfy the Penrose property.
\end{prop}
\textbf{Proof:} The compactified Schwarzschild line element (\ref{eqn:compactified}) differs from the compactified Minkowski metric (\ref{eqn:compactMink}) only by the function $\frac{r^2}{V(r)r_*^2}$ which appears in front of the angular terms (this function is identically $1$ for Minkowski). In fact, the similarity is deeper than this. Not only do the metrics look similar, we also note that the negative mass Schwarzschild spaectime is mapped to the same compactified manifold as the $r>0$ portion of Minkowski. This means we have a bijection between curves in compactified Schwarzschild and curves in compactified Minkowski (with $r>0$) which arises from identifying the co-ordinates $(T,\chi,\theta,\phi)$ defined on both these manifolds. It does not matter that these will correspond to different curves when we map back to the physically relevant uncompactified spacetimes.

To further this comparison, we need to understand the relation between the compactified metrics, which means looking more closely at the function $\frac{r^2}{V(r)r_*^2}$. We consider the large $r$ asymptotics of this function. Expanding $V(r)$ and $r_*(r)$ to leading order, we have
\begin{equation}\label{eqn:larger}
    \begin{split}
       \frac{r^2}{r_*^2V(r)}&=\left[1+\frac{2r_s}{r}\log\left(r/r_s+1\right)+O\left(\frac{r_s^2}{r^2}\left(\log\left(r/r_s+1\right)\right)^2\right)\right]\left[1-\frac{r_s}{r}+O\left(\frac{r_s^2}{r^2}\right)\right]\\
        &=1+\left(2\log\left(r/r_s+1\right)-1\right)\frac{r_s}{r}+O\left(\frac{r^2}{r^2_s}\left(\log\left(r/r_s+1\right)\right)^2\right)\\
        &>1 \text{ for large }r
    \end{split}
\end{equation}
This means the conformally compactified negative mass Schwarzschild metric in 3+1 dimensions can be bounded above by the conformally compactified Minkowski metric, i.e. $g<\eta$ (for large $r$).

We conclude from this that the Penrose property does not hold for negative mass in 3+1 dimensional Schwarzschild spacetime. In particular, if we choose antipodal points on $\mathcal{I}^-$ and $\mathcal{I}^+$ sufficiently near $i^0$, then these cannot be timelike connected. To see this, suppose we could connect such points using a Schwarzschild timelike curve. By choosing these points sufficiently close to $i^0$, we can ensure that such a curve is restricted to large enough $r$ to ensure that inequality (\ref{eqn:larger}) holds. This is because curves near $i^0$ in compactified Schwarzschild are necessarily restricted to large values of $r_*$ (since $(T,\chi)=(0,\pi)$ at $i^0$ with these co-ordinates defined as in equation (\ref{eqn:Tchi})) and from equation (\ref{eqn:negativemasstortoise}) we see that $r\rightarrow\infty$ as $r_*\rightarrow\infty$. This curve would then also be timelike in compactified Minkowski spacetime and would connect the same points on $\mathcal{I}^\pm$. But we know (section \ref{Minkowski Causality}) that antipodal points sufficiently near spatial infinity in Minkowski spacetime cannot be timelike connected, so we conclude (choosing our points even closer to $i^0$ if necessary) that the Penrose property does not hold for negative mass Schwarzschild in 3+1 dimensions. \qed

This argument is similar to the argument used for negative mass in 2+1 dimensions. The difference is that the constant $\alpha\in(0,1)$ which created the deficit angle has now been replaced by the function $\frac{r^2}{V(r)r_*^2}\in(0,1)$ which depends on $r$.
\subsection{Schwarzschild Spacetime in $d+1$ dimensions with $d>3$}
\subsubsection{Positive Mass}
\label{Positive Mass Schwarzschild in Higher Dimensions}
It is tempting to try to use the same construction as \cite{Penrose} (and summarised in section \ref{Positive Mass Schwarzschild in 3+1 Dimensions}) to show that the Penrose property is true for positive mass Schwarzschild in 3+1 dimensions. In fact, this construction no longer works because the endpoints of non-radial null geodesics do not approach $i^0$ as we let the impact parameter $R\rightarrow\infty$. In fact, the situation for positive mass Schwarzschild in higher dimensions is similar to the negative mass case in 3+1 dimensions as we see below.
\begin{prop}
\label{prop71}
Positive mass Schwarzschild spacetime in $d+1$ dimensions ($d>3$) does not satisfy the Penrose property.
\end{prop}
\textbf{Proof:} The Schwarzschild metric in $d+1$ dimensions $(d>3)$ is
\begin{equation}
\label{eqn:schwarzschildhigher}
    \begin{split}
        ds^2=V(r)dt^2-\frac{dr^2}{V(r)}-r^2d\omega_{d-1}^2
    \end{split}
\end{equation}
where $V(r)=1-\frac{r_s^n}{r^n}$, $n=d-2$ and $d\omega_{d-1}^2$ is the round metric on $S^{d-1}$.

We can conformally compactify the metric as in lower dimensions to obtain the same form as equation (\ref{eqn:compactified}) (with $d\omega^2_2$ replaced by $d\omega^2_{d-1}$ throughout).

Once again we consider the large $r$ asymptotics of the metric. We have
\begin{equation}\label{eqn:inequality}
    \begin{split}
        dr_*&=\frac{dr}{V(r)}=\left(1-\frac{{r_s}^n}{r^n}\right)^{-1}dr
=\left(1+\frac{{r_s}^n}{r^n}+O\left(\frac{{r_s}^{2n}}{r^{2n}}\right)\right)dr\\
        \implies r_*&=r\left(1-\frac{{r_s}^n}{(n-1)r^{n}}+O\left(\frac{{r_s}^{2n}}{r^{2n}}\right)\right)
\\
        \implies \frac{r^2}{r_*^2V(r)}&=\left(1+\frac{2r_s^n}{(n-1)r^{n}}+O\left(\frac{r_s^{2n}}{r^{2n}}\right)\right)\left(1+\frac{r_s^n}{r^n}+O\left(\frac{r_s^{2n}}{r^{2n}}\right)\right)\\
        &=1+\frac{(n+1)r_s^n}{(n-1)r^n}+O\left(\frac{r_s^{2n}}{r^{2n}}\right)\\
        &>1\text{ for large }r.
    \end{split}
\end{equation}
So we obtain the same bounds as in section \ref{Negative Mass Schwarzschild in 3+1 Dimensions}, and the same argument tells us that the Penrose property does not hold in this spacetime (again noting that curves near $i^0$ in Schwarzschild are restricted to large values of $r_*$ and that $r\rightarrow\infty$ as $r_*\rightarrow\infty$).$\qed$
\subsubsection{Negative Mass}
\label{Negative Mass Schwarzschild in Higher Dimensions}
The negative mass Schwarzschild metric in spacetime of dimension $d+1$ ($d>3$) is as in equation (\ref{eqn:schwarzschildhigher})
except with $V(r)=1+\frac{r_s^n}{r^n}$, ($n=d-2$). The bound we obtained in (\ref{eqn:inequality}) is now reversed, so we cannot use this method to conclude that the Penrose property is not satisfied. In fact, the lightcones of the naturally associated Minkowski metric are now contained inside the Schwarzschild lightcones (at least at large $r$). This suggests that the spacetime may have a chance of satisfying the Penrose property (we should at least be able to timelike connect all the same points as we could in Minkowski by using curves restricted to large $r$ where this bound holds). However, as we let $r\rightarrow\infty$, the Schwarzschild and Minkowski lightcones converge towards each other. This means that we may be unable to connect antipodal points sufficiently near $i^0$, as is the case in Minkowski spacetime, since such curves would be restricted to increasingly large values of $r$. In fact, it turns out that the deflection of null geodesics away from the negative point mass (see Figure \ref{fig:positivemass}) prevents us from timelike connecting certain pairs of antipodal points near $i^0$.
\begin{thm} \label{thm:NegativeMass}
Negative mass Schwarzschild spacetime in $d+1$ dimensions ($d>3$) does not satisfy the Penrose property.
\end{thm}
To prove this theorem, we will require the following result:
\begin{thm}\label{thm:Seifert}\textbf{(Seifert \cite{Seifert}, Thm. 1)}
Let $a,b\in M$ be such that the causal diamond $J^+(a)\cap J^-(b)$ is non-empty, compact and strongly causal. Then there exists a causal geodesic from $a$ to $b$.
\end{thm}
\textbf{Proof of Theorem \ref{thm:NegativeMass}:} Suppose we are aiming to timelike connect points $p\in\mathcal{I}^-$ and $q\in\mathcal{I}^+$. We choose co-ordinates $(t,r,\theta_1,...,\theta_{d-2},\phi)$ such that $\theta_1=...=\theta_{d-2}=\pi/2$ at $p$ and $q$. Once again we can exploit the staticity and spherical symmetry of the metric and argue using Lemma \ref{spherical symmetry} that it suffices to consider curves restricted to the surface $\theta_1=...=\theta_{d-2}=\pi/2$. All curves and points referred to in this proof will be assumed to lie on this surface.
\begin{figure} 
  \minipage{0.48\textwidth}
  \centering
        \includegraphics[scale=0.3]{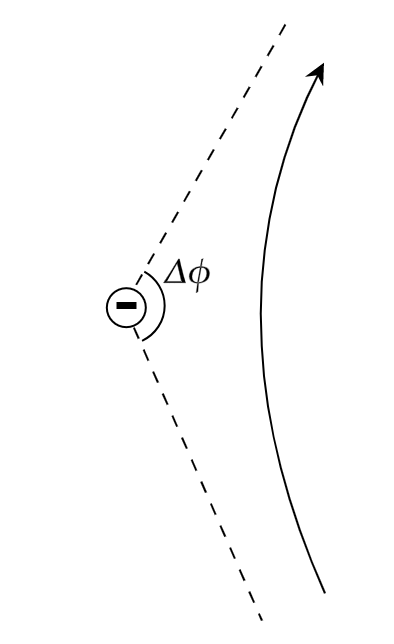}
    \caption{$|\Delta\phi|<\pi$ along null geodesics in negative mass Schwarzschild.}
    \medskip
    \medskip\medskip
    \label{fig:negativemass}
    \endminipage
     \minipage{0.04\textwidth}
      \includegraphics[scale=0.05]{blank.png}
      \endminipage
    \minipage{0.48\textwidth}
 \centering
    \includegraphics[scale=0.305]{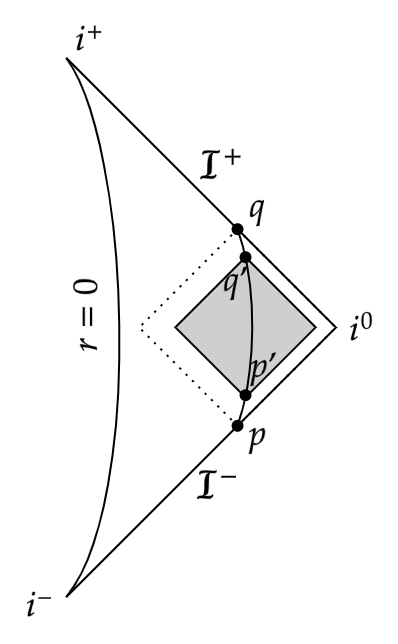}
    \caption{The causal diamond (shaded) formed by the points $p'$ and $q'$ is non-empty, compact and strongly causal, so we can apply Lemma \ref{thm:Seifert}.}
    \label{fig:causaldiamond}
    \endminipage
\end{figure}
A negative point mass will be repulsive and will deflect null geodesics away (Figure \ref{fig:negativemass}). We therefore expect that, in contrast to positive mass, causal geodesics will undergo an angular change of strictly less than $\pi$. This is what we now prove. Solving the geodesic equations for the Schwarzschild metric, we find
\begin{equation}
    \begin{split}
    r^2\dot{\phi}&=h=\text{constant}\\
        \text{and }\dot{r}^2&=1-V(r)\left(k^2+\frac{h^2}{r^2}\right)\\
    \end{split}
\end{equation}
where $k\geq0$ is a constant such that $k=0$ for a null geodesic and $k>0$ for a timelike geodesic. We can determine $h^2$ in terms of $R$ (the minimum value of $r$ along the geodesic) by solving $\dot{r}=0$ at $r=R$. We find that
\begin{equation}\label{eqn:ineq3}
\begin{split}
    h^2&=\frac{R^2}{V(R)}\left(1-k^2V(R)\right)\\
    \implies\frac{\dot{r}^2}{h^2}&=\frac{V(R)}{R^2}\frac{1-k^2V(r)}{1-k^2V(R)}-\frac{V(r)}{r^2}\\
    &\geq\frac{V(R)}{R^2}-\frac{V(r)}{r^2}\\
    &\geq\frac{V(R)}{R^2}\left(1-\frac{R^2}{r^2}\right)
    \end{split}
\end{equation}
where we have used the the fact that $V(r)\leq V(R)$ for $r\geq R$ and we also assume that $h\neq0$ in order to make the inequalities easier to derive (the case $h=0$ is trivial). This gives an upper bound for the magnitude of the change in $\phi$ along a causal geodesic, denoted $|\Delta\phi|$:
\begin{equation}\label{eqn:deltaphi2}
    \begin{split}
        |\Delta\phi|&=2|h|\int_R^\infty\frac{dr}{\dot{r}r^2}\\
        &\leq\frac{2R}{V(R)^{1/2}}\int_R^\infty\frac{dr}{(1-R^2/r^2)^{1/2}r^2}\\
        &=\frac{\pi}{V(R)^{1/2}}\\
        &<\pi
    \end{split}
\end{equation}
where the final inequality is obtained from the fact that $V(R)=1+\frac{r_s^n}{R^n}>1$. Note that the first inequality in (\ref{eqn:ineq3}) tells us that, as expected, among all causal geodesics with fixed $h$, $|\Delta\phi|$ is maximised by null geodesics (for which $k=0$).

We aim to apply Theorem \ref{thm:Seifert} to the causal diamond formed by two points in the uncompactified spacetime (since inequality (\ref{eqn:deltaphi2}) refers to geodesics in this spacetime). Besides, since $i^0$ is singular, we could not apply the Theorem to points on null infinity.  In order to apply Theorem \ref{thm:Seifert}, we need to make sure that the causal diamond does not intersect the naked singularity at $r=0$.  

Since the Penrose property relates to the compactified spacetime, we
begin by choosing $p\in\mathcal{I^-}$ and $q\in\mathcal{I}^+$ sufficently
close to $i^0$ that $J^+(p)\cap J^-(q)$ does not intersect the $r=0$
singularity. We can do this because the compactified spacetime can be
decomposed as $\Sigma\times S^{d-1}$ as in Lemma \ref{spherical symmetry} . The causal diamond formed by any $p\in\mathcal{I}^-$ and $q\in\mathcal{I}^+$ then consists of a region on the $(d-1)$-sphere and a causal diamond on $\Sigma$ (as
shown by the dotted line in Figure  \ref{fig:causaldiamond} ), which can certainly be chosen so as to avoid the $r=0$ singularity.

Note in particular that the causal diamond formed by any two points in $J^+(p)\cap J^-(q)$ will also not intersect this singularity. Now if the Penrose property were satisfied in this spacetime then there would exist a timelike curve from $p$ to $q$ (recall that this curve would be timelike with respect to both the compactified and uncompactified metrics). Since the metric is invariant under $\phi\mapsto-\phi$, we can assume without loss of generality that $\phi$ is increasing along this curve. Now for any point $a\in\tilde{M}$, $I^\pm(a)$ are open sets (\cite{DiffTop}, Prop. 2.8), so we must be able to modify this curve to obtain a new timelike curve, with the same $u$ and $v$ values at its endpoints, except with $\phi=-\epsilon$ on $\mathcal{I}^-$ and $\phi=\pi+\epsilon$ on $\mathcal{I}^+$ (for some $\epsilon>0$) and such that $\phi$ is still increasing along this new curve. By restricting to a segment of this curve, we obtain a timelike curve between points $p'$ and $q'$ in the interior spacetime, where $\phi=0$ at $p'$ and $\phi=\pi$ at $q'$. The set $J^+(p')\cap J^-(q')$ (or more precisely, $\lambda^*\left(J^+(p')\cap J^-(q')\right)$, the pullback to the uncompactified manifold of this set by the embedding map defined in Definition \ref{defn:asymptotically empty and simple}) then satisfies the conditions of Theorem \ref{thm:Seifert} (in particular it is free of any singular points). We therefore conclude that there must exist a causal geodesic (with respect to the uncompactified Schwarzschild metric) from $\lambda^*(p')$ to $\lambda^*(q')$. But we also know that such a geodesic cannot exist. In particular, it would necessarily have $\Delta\phi\geq\pi$ along its full length which would contradict inequality (\ref{eqn:deltaphi2}). We therefore conclude that the Penrose property is not satisfied by negative mass Schwarzschild in higher dimensions. \qed

Note that the spacetime is singular at $i^0$, so it was necessary to choose $p'$ and $q'$ to lie in the interior of the spacetime in order to ensure that $i^0\notin J^+(p')\cap J^-(q)$ so we are able to apply Theorem \ref{thm:Seifert} to this region. 

\subsection{Summary of the Penrose Property}\label{Summary of the Penrose Property}
Propositions \ref{prop41}, \ref{prop61}, \ref{prop71}, \ref{thm:NegativeMass} and Theorem \ref{thm:51} together establish Theorem A from the introduction which we now re-state.

\textbf{Theorem A:}
{\em
The Penrose property is satisfied by Schwarzschild spacetime of mass $m$ and varying spacetime dimension according to the following table
 \begin{center}
    \begin{tabular}{c|c|c}
  Spacetime dimension & $m>0$ & $m\leq0$\\
   \hline
 $3$ & \cmark & \xmark\\ 
 $4$ & \cmark & \xmark\\  
 $\geq5$ & \xmark &  \xmark
    \end{tabular}
\end{center}
}
The Penrose property appears to be a property of positive mass in low dimensions. We have seen that adding positive mass in 3 or 4 dimensions gives us the inequality
\begin{equation}
    \label{eqn:conesineq}
    \tilde{ds}^2_S\geq\tilde{ds}^2_M
\end{equation}
at large $r$, where $\tilde{ds}^2_M$ is the naturally associated compactified Minkowski metric. This tells us that the lightcone condition $g<\eta$ fails.  This is also the case for negative mass in higher dimensions (although in this case, rather than diverging, the cones converge together at infinity and, as we have seen, the effect of the mass is not strong enough for the Penrose property to be satisfied). For negative mass in 3 and 4 dimensions as well as positive mass in higher dimensions, the inequality (\ref{eqn:conesineq}) is now reversed at large $r$ and the Penrose property fails (by comparison with Minkowski space). It is curious that positive and negative mass seem to have the opposite effect on lightcones in higher dimensions compared with 3 and 4 dimensions.

Since the Penrose property is concerned only with the asymptotic behaviour of the metric near spatial infinity, these results can be immediately applied to a number of other spacetimes. For example the Reissner-Nordstr\"om line element in $d+1$ dimensions ($d\geq3$) is
$$ds^2=\left(1-\frac{2m}{r^{d-2}}+\frac{Q^2}{r^{2d-4}}\right)dt^2-\left(1-\frac{2m}{r^{d-2}}+\frac{Q^2}{r^{2d-4}}\right)^{-1}dr^2-r^2d\omega_{d-1}^2$$
where $Q$ is the charge of the black hole. Just like Schwarzschild, this spacetime satisfies the Penrose property only when $d=3$ and $m>0$. The case where $m=0$, $Q\neq0$ is notable because the large $r$ asymptotics are now the same as those of negative mass Schwarzschild in $2d-2$ spatial dimensions.
\section{Kerr Spacetime}\label{Kerr Metric}
We can also use this method of comparing metrics to gain insight into the Penrose property of a spacetime which is not spherically symmetric, namely the Kerr spacetime.

The Kerr metric in Boyer-Lindquist co-ordinates is
$$ds^2_{K,m}=\left(1-\frac{2mr}{\Sigma}\right)dt^2-\frac{\Sigma}{\Delta_m}dr^2-\Sigma d\theta^2-\Lambda_m^2\sin^2\theta d\phi^2+\frac{4mar\sin^2\theta}{\Sigma}d\phi dt\\$$
where we define
\begin{equation}
    \begin{split}
        \Delta_m(r)&= r^2-2mr+a^2\\
        \Sigma(r,\theta)&=r^2+a^2\cos^2\theta\\
        \Lambda^2_m(r,\theta)&=r^2+a^2+\frac{2ma^2r\sin^2\theta}{\Sigma}
    \end{split}
\end{equation}
and $a=\frac{J}{m}$ is a constant rotation parameter written in terms of the angular momentum $J$ and the mass $m>0$ of the black hole. The subscripts $m$ refer to the mass parameter and have been inserted for later convenience.

\textbf{Theorem B:}
{\em The (positive mass) Kerr spacetime in $3+1$ dimensions satisfies the Penrose property.
}

To prove this theorem, it will be useful to refer to the following metric which will be used to bound the Kerr line element.
\begin{definition}
For any $R>2m>0$, we define the \textit{quasi-Schwarzschild} metric with parameters $(m,a)$ by
$$ds^2_{QS,m}=\left(1-\frac{2m}{r}\right)dt^2- \frac{dr^2}{1-\frac{2m}{r}}-r^2d\theta^2-r^2\sin^2\theta\left(d\phi-\omega_{m,\alpha}\left(r,\theta\right) dt\right)^2$$
where $\alpha^2:=1+\frac{a^2}{R^2}\left(1+\frac{2m}{R}\right)$, $\omega_{m,\alpha}( r,\theta):=\frac{2mar\alpha^{-2}}{\Sigma(\frac{r}{\alpha},\theta) \Lambda^2_\frac{m}{\alpha}(\frac{r}{\alpha},\theta)}$. Again we have inserted subscripts for later convenience.
\end{definition}

This metric is similar to the $3+1$ dimensional Schwarzschild metric of mass $m$, the only difference being the $\omega dt$ term in the final bracket. Since we are interested in neighbourhoods of spatial infinity, it suffices to define the quasi-Schwarzschild metric on the manifold $M_R:=\mathbbm{R}\times\left(\mathbbm{R}^3\backslash B_R(0)\right)$, where $B_R(0)$ denotes the ball of radius $R$ centred at the origin, for some $R>2m$. For $r\geq R$, the components of the quasi-Schwarzschild metric are then smooth functions of the co-ordinates. 

We can now prove the following lemma, which will be used in the proof of Theorem B.
\begin{lemma}\label{lemma:Kerrbound}
For $r\geq R$, the line element of the Kerr metric with mass $m$ and rotation parameter $a$ can be bounded from below by a quasi-Schwarzschild line element with parameters $(\tilde{m}=\alpha m,a)$:
$$ds^2_{K,m}\geq ds^2_{QS,\tilde{m}}$$
\end{lemma}
\textbf{Proof:}
\begin{equation}
    \begin{split}
        ds^2_{K,m}    =&\left(1-\frac{2mr}{\Sigma}+\frac{4m^2a^2r^2\sin^2\theta}{\Sigma^2\Lambda_m^2}\right)dt^2-\frac{\Sigma}{\Delta_m}dr^2-\Sigma d\theta^2-\Lambda_m^2\sin^2\theta\left(d\phi-\frac{2mar}{\Sigma \Lambda_m^2}dt\right)^2\\
        \geq& \left(1-\frac{2m}{r}\right)dt^2
        -\frac{r^2+a^2}{r^2-2mr}dr^2-(r^2+a^2)d\theta^2\\
        &-\left(r^2+a^2\left(1+\frac{2m}{r}\right)\right)\sin^2\theta\left(d\phi-\omega_{\alpha m,\alpha}(\alpha r, \theta) dt\right)^2\\
        \geq&\left(1-\frac{2m}{r}\right)dt^2- \frac{\alpha^2dr^2}{1-\frac{2m}{r}}-\alpha^2r^2d\theta^2-\alpha^2r^2\sin^2\theta\left(d\phi-\omega_{\alpha m,\alpha}(\alpha r, \theta) dt\right)^2\\
        =&\left(1-\frac{2\Tilde{m}}{\Tilde{r}}\right)dt^2- \frac{d\Tilde{r}^2}{1-\frac{2\Tilde{m}}{\Tilde{r}}}-\Tilde{r}^2d\theta^2-\Tilde{r}^2\sin^2\theta\left(d\phi-\omega_{\tilde{m},\alpha}(\tilde{r}, \theta) dt\right)^2\\
        =&ds^2_{QS,\tilde{m}}
    \end{split}
\end{equation}
where we have defined the new variable $\tilde{r}=\alpha r$.\qed

From now on we will omit the subscripts $\alpha$, $m$ and $\tilde{m}$, with the understanding that for $r>R$, the Kerr spacetime of mass $m$ should be compared to the quasi-Schwarzschild spacetime of mass $\tilde{m}=\alpha m$. The actual numerical value of $\tilde{m}>0$ is unimportant. We simply require the existence of some quasi-Schwarzschild metric with positive mass with which to bound the Kerr metric. 

\textbf{Proof of Theorem B:} We begin by compactifying the Kerr and quasi-Schwarzschild spacetimes. To compactify the quasi-Schwarzschild spacetime, we can define the same compactified co-ordinates and conformal factor as were used for Schwarzschild in section \ref{Positive Mass Schwarzschild in 3+1 Dimensions}. This gives an embedding
$$\lambda_{QS}:(M_R,g_{QS})\rightarrow(\tilde{M}_R,\tilde{g}_{QS})$$
from the quasi-Schwarzschild spacetime to its conformal compactification. The map $\lambda_{QS}$ is invertible, so it is possible to express the $\omega_{m,\alpha}(r,\theta)dt$ term of the quasi-Schwarzschild line element in terms of the compactified co-ordinates. 

The compactification of the Kerr metric can be carried out by defining Kruskal co-ordinates as in Section 8 of \cite{Kerr}. This gives an embedding
$$\lambda_K:(M_R,g_K)\rightarrow(\tilde{M}_R\tilde{g}_K)$$
We will use the fact that the compactifications of the Schwarzschild, quasi-Schwarzschild and Kerr metrics can all be defined on the same manifold $\tilde{M}_R$, so in particular we can identify points on $\mathcal{I}^\pm$ in these three spacetimes. This will be important in the argument below.

The compactified Kerr and compactified quasi-Schwarzschild line elements are conformal to the uncompactified versions. We have
$$\tilde{ds}^2_K=\Omega_{K}^2 ds^2_K$$
$$\tilde{ds}^2_{QS}=\Omega_{QS}^2 ds^2_{QS}$$
for some functions $\Omega_{QS}$ and $\Omega_{K}$ which in particular are strictly positive away from the boundary of the compactified spacetimes. From Lemma \ref{lemma:Kerrbound} we have, for $r>R$:
\begin{equation}\label{eqn:ineq}
    \begin{split}
    ds^2_K&\geq ds^2_{QS}\\
       \implies\Omega_{K}^{-2}\Tilde{ds}_K^2&\geq\Omega^{-2}_{QS} \Tilde{ds}^2_{QS}\\
       \implies\Tilde{ds}_{K}^2&\geq\frac{\Omega_{K}^2}{\Omega^2_{QS}} \Tilde{ds}^2_{QS}
    \end{split}
\end{equation}

So we have a lower bound for the compactified Kerr line element in terms of the compactified quasi-Schwarzschild line element. This bound tells us that the mapping $\lambda_K\circ\lambda_{QS}^{-1}$ maps timelike curves in compactified quasi-Schwarzschild spacetime to timelike curves in compactified Kerr spacetime.

Next we investigate how points on $\mathcal{I}^\pm$ behave under this mapping. It is possible that they relate to entirely different points on $\mathcal{I}^\pm$ \footnote{This is precisely why this argument fails to show, for example, that positive mass Schwarzschild spacetime does not satisfy the Penrose property. In this situation we have $ds^2_S\leq ds^2_M$, however when we compactify both metrics, the compactified co-ordinates diverge away from each other as we approach null infinity. This means that all timelike curves in compactified Schwarzschild correspond to curves between past and future timelike infinity in compactified Minkowski. As a result, we cannot use the failure of the Penrose property in Minkowski spacetime to deduce that it also fails in positive mass Schwarzschild.}. However, suppose we choose compactified co-ordinates defined by ``quasi-spherical light cones" as described in section 5 of \cite{Kerr}. This means compactifying the metric as in section \ref{Positive Mass Schwarzschild in 3+1 Dimensions}, with $r_*$ defined as in \cite{Kerr}. Then these compactified co-ordinates agree with the compactified Schwarzschild co-ordinates from section \ref{Positive Mass Schwarzschild in 3+1 Dimensions} as $r\rightarrow\infty$, so points on $\mathcal{I}^\pm$ are fixed under the mapping $\lambda_K\circ\lambda_{QS}^{-1}$. In particular, this means that if we have a timelike curve between two points on $\mathcal{I}^\pm$ in compactified quasi-Schwarzschild spacetime, then under the mapping $\lambda_K\circ\lambda_{QS}^{-1}$, this becomes a timelike curve between the same two points on $\mathcal{I}^\pm$ in the compactified Kerr spacetime. 

Hence in order to show that the Penrose property holds in Kerr spacetime, it suffices to show that it holds in quasi-Schwarzschild spacetime. This is what we will now prove. 
\begin{figure} 
    \centering
    \includegraphics[scale=0.3]{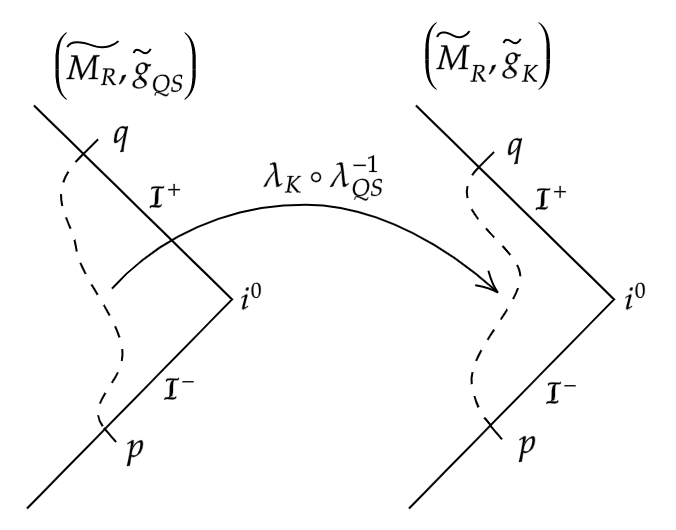}
    \caption{The function $\lambda_K\circ\lambda_{QS}^{-1}$ maps timelike curves in $(\tilde{M}_R,\tilde{g}_{QS})$ to timelike curves in $(\tilde{M}_R,\tilde{g}_K)$. This mapping also preserves the endpoints of these curves on $\mathcal{I}^\pm$.}
    \label{fig:QStoKerr}
\end{figure}
 For this argument, we will refer to the Schwarzschild spacetime $(M_R,g_S)$ in 3+1 dimensions with mass $m>0$. Suppose we wish to find a timelike curve in compactified quasi-Schwarzschild spacetime connecting the point $p\in\mathcal{I}^-$ with co-ordinates $(T,\chi,\theta,\phi)=(T_0,\chi_0,\theta_0,0)$ to the point $q\in\mathcal{I}^+$ with co-ordinates $(T_1,\chi_1,\theta_1,\phi_1)$ (we have exploited the axial symmetry of $(\tilde{M}_R,\tilde{g}_{QS})$ to set $\phi=0$ at $p$). Since $(M_R,g_S)$ satisfies the Penrose property, we can find a curve, $\gamma_S$, which is timelike with respect to $g_S$ and which connects the point $p$ to the point on $\mathcal{I}^+$ with co-ordinates $(T_1,\chi_1,\theta_1,\pi)$.

We can parametrize this curve by $t$ and write the compactified co-ordinates along the curve as functions $T=T(t)$, $\chi=\chi(t)$, $\phi=\phi_S(t)$, $\theta=\theta(t)$. We have used a subscript $S$ on the $\phi$ co-ordinate function to note that this refers to the Schwarzschild timelike curve $\gamma_S$. This is the function we will change when we construct a curve which is timelike with respect to the quasi-Schwarzschild metric.

We can also write $\omega=\omega(r(t),\theta(t))$ along this curve. This allows us to compute the following quantity:
$$A:=\int_{t=-\infty}^{t=\infty}\omega(r(t),\theta(t))dt$$
Note that this integral converges since $\omega$ is finite at any $r$, $\theta$ (assuming $a\neq0$) and for a path with endpoints on $\mathcal{I}^\pm$ we have 
$$\omega\sim\frac{1}{r^3}\sim\frac{1}{r_*^3}\sim\frac{1}{|t|^3}\text{ as }|t|\rightarrow\infty$$
where $r_*$ is the co-ordinate defined in Section 5 of \cite{Kerr}. The point about this definition is that
$t\pm r_*$ tends to a constant near $\mathcal{I^\pm}$ and the leading order difference between $r_*$ and $r$ is a term which is logarithmic in $r$.

Observe that $A$ is independent of the function $\phi_S(t)$. It depends only on the path in $(T,\chi,\theta)$ space (since $(t,r)$ are defined only in terms of $(T,\chi)$).

Next note that if a curve is timelike in compactified Schwarzschild it will still be timelike if we reduce (and possibly reverse) the rate at which the $\phi$ co-ordinate varies, while keeping the same path in $(T,\chi,\theta)$ space. By adding a suitable number of integer multiples of $2\pi$ to $A$ if necessary, we may assume that $\phi_1-A\in(-\pi,\pi]$. We then define a new curve $$\hat{\gamma}_S(t)=(T(t),\chi(t),\frac{1}{\pi}\left(\phi_1-A\right)\phi_S(t),\theta(t))$$

This curve is the same as $\gamma_S$ except with the $\phi$ co-ordinate function re-scaled by the constant $\frac{1}{\pi}(\phi_1-A)\in(-1,1]$, which in particular means that $\hat{\gamma}_S$ is timelike. It is important to note that $A$ remains unchanged by this re-scaling. The reason for this re-scaling is that the endpoints of the curve $\hat{\gamma}_S$ satisfy
\begin{equation}
    \begin{split}
        \hat{\phi}_S(t=-\infty)&=0\\
\hat{\phi}_S(t=\infty)&=\phi_1-A
    \end{split}
\end{equation}

We therefore have a curve, $\hat{\gamma}_S$, which is timelike with respect to $g_S$ and which connects $p\in\mathcal{I}^-$ to the point $q'\in\mathcal{I}^+$ with co-ordinates $(T_1,\chi_1,\phi_1-A,\theta_1)$.

Using this we can construct a curve, $\gamma_{QS}$, from $p$ to $q$ which is timelike with respect to $g_{QS}$. To do this, we define $\gamma_{QS}$ to follow the same path in $(T,\chi,\theta)$ space as the curve $\hat{\gamma}_S$. The $\phi$ co-ordinate along $\gamma_{QS}$ will be given by a function, $\phi_{QS}(t)$, defined such that
\begin{equation}\label{eqn:curves}
    \Tilde{ds}_{QS}(t)\equiv\tilde{ds}_S(t)\geq0
\end{equation}
 at every point on $\gamma_{QS}$ and $\hat{\gamma}_{S}$. This ensures that $\gamma_{QS}$ is timelike with respect to the quasi-Schwarzschild metric $g_{QS}$.
\begin{figure} 
    \centering
    \includegraphics[scale=0.3]{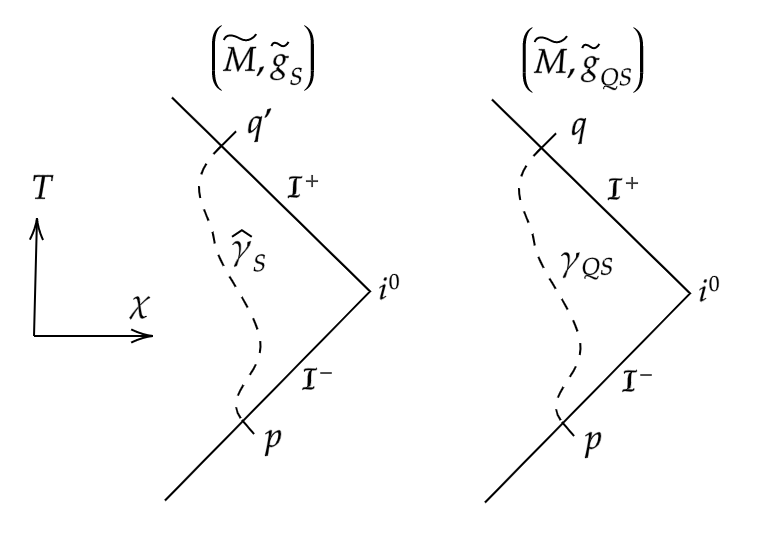}
    \caption{$\hat{\gamma}_S$ is a timelike curve in $(\tilde{M}_R,\tilde{g}_S)$ between $p$ and $q'$. Using this curve, we define a curve $\gamma_{QS}$ which is timelike in $(\tilde{M}_R,\tilde{g}_{QS})$ and which connects $p$ and $q$. These curves follow the same path in $(T,\chi,\theta)$ space.}
    \label{fig:StoQS}
\end{figure}
Since $\hat{\gamma}_S$ and $\gamma_{QS}$ follow the same path in $(T,\chi,\theta)$ space (so the functions $r(t)$, $\theta(t)$ are the same along both curves), equation (\ref{eqn:curves}) becomes the requirement that 
$$\dot{\phi}_{QS}(t)=\dot{\hat{\phi}}_S(t)+\omega(r(t),\theta(t))$$
To satisfy this, we define the function $\phi_{QS}(t)$ by
$$\phi_{QS}(t)=\hat{\phi}_S(t)+\int_{-\infty}^t\omega(r(t'),\theta(t'))dt'$$
This means that at the future endpoint of $\gamma_{QS}$, the $\phi$ co-ordinate is 
$$\phi_{QS}(t=\infty)=\int_{-\infty}^{\infty}\dot{\phi}_{QS}(t)dt=\int_{-\infty}^{\infty}\left[\dot{\hat{\phi}}_S(t)+\omega(r(t),\theta(t))\right]dt=\phi_0-A+A=\phi_0$$
as required.$\qed$
\section{Unique Continuation of the Linear Wave Equation}\label{uniquecontinuation}
\subsection{The Null Geodesic Endpoint Condition}
The Penrose property, or more specifically the property of null geodesic endpoints proved by Penrose for positive mass Schwarzschild in $3+1$ dimensions \cite{Penrose}, can help with understanding the problem of unique continuation from null infinity of the linear wave equation. In \cite{LinearWaves}, the authors consider the linear wave equation
\begin{equation}\label{eqn:wave}
    \Box_g\varphi+a^\mu\partial_\mu\varphi+V\varphi=0
\end{equation}
where $\Box_g$ is the Laplace-Beltrami operator for the metric $g$ and the functions $a^\mu, V$ satisfy certain fall-off conditions (see \cite{LinearWaves}). They investigate conditions on the solution at null infinity which ensure that it can be uniquely continued into some region of the interior spacetime. 

Two separate classes of spacetimes are considered. The first is Minkowski spacetime in $d+1$ dimensions ($d\geq2$) along with perturbations to the metric described in \cite{LinearWaves} (which are such that the ADM mass \cite{ADM} of the spacetime remains zero). The second class is Schwarzschild spacetime in $3+1$ dimensions along with metric perturbations \cite{LinearWaves}. Again these perturbations are sub-leading in the sense that they do not change the ADM mass of the spacetime. Note that this set-up is consistent with the proof in \cite{Penrose}, summarised in section \ref{Positive Mass Schwarzschild in 3+1 Dimensions}, that we cannot think of Schwarzschild spacetime in 3+1 dimensions as arising from a perturbation of Minkowski due to differences in their asymptotic behaviours near $i^0$.

Following \cite{LinearWaves}, we define the following subsets of null infinity
\begin{equation}\label{eqn:intervals}
    \mathcal{I}_{u_0}^+=\{v=\infty, \textbf{ }u\leq u_0\},\textbf{ }\mathcal{I}_{v_0}^-=\{u=-\infty, \textbf{ }v\geq-v_0\}
\end{equation}
for any $u_0,v_0\in\mathbbm{R}$, where the retarded and advanced time co-ordinates $u$ and $v$ are given by equation (\ref{eqn:MinkRet}) for perturbations of the Minkowski metric and equation (\ref{eqn:SchwRet}) for perturbations of the Schwarzschild metric. For some $u_0,v_0\in\mathbbm{R}$, we will be interested in uniquely continuing the solution to equation (\ref{eqn:wave}) from $\mathcal{I}_{u_0}^+\cup\mathcal{I}_{v_0}^-$ into a  region
\begin{equation}\label{eqn:region}
    \mathcal{D}^{(u_0, v_0)}_\omega:=\left\{0<\frac{1}{(v+v_0)(u_0-u)}<\omega\right\}
\end{equation}
where $\omega>0$.

It turns out that for perturbations of the Minkowski metric, if we impose conditions on the solution on subsets $\mathcal{I}^+_{\epsilon}\subset\mathcal{I}^+$ and $\mathcal{I}^-_{\epsilon}\subset\mathcal{I}^-$ for any $\epsilon>0$, then this solution can be uniquely continued into the interior spacetime . In particular, if we demand that $\varphi$ and its first derivatives vanish to infinite order on $\mathcal{I}_\epsilon^+\cup\mathcal{I}^-_\epsilon$ (for $\epsilon>0$), then $\varphi$ must vanish on some open domain in the interior spacetime which contains $\mathcal{I}_\epsilon^+\cup\mathcal{I}^-_\epsilon$ on its boundary, i.e. $\varphi=0$ on $\mathcal{D}_\omega^{(\epsilon,\epsilon)}$ for some $\omega>0$

Now consider perturbations of the Schwarzschild metric in $3+1$ dimensions (which in particular include the Kerr spacetime). We define the sets $\mathcal{I}^+_{u_0}$, $\mathcal{I}_{v_0}^-$ and $ \mathcal{D}^{(u_0, v_0)}_\omega$ as in (\ref{eqn:intervals}) and (\ref{eqn:region}), with the retarded and advanced time co-ordinates now defined by
$$u=t-r-r_s\log(r/r_s-1)$$
$$v=t+r+r_s\log(r/r_s-1)$$
Suppose we demand that the solution, $\varphi$, and its first derivatives vanish to infinite order on 
$\mathcal{I}_{u_0}^+\cup\mathcal{I}_{v_0}^-$
for any $u_0$, $v_0\in\mathbbm{R}$. Then in \cite{LinearWaves} it is shown that $\varphi$ must vanish in $\mathcal{D}_\omega^{(u_0,v_0)}$ for some $\omega>0$.

Note that, in contrast to the case of perturbations of the Minkowski metric, we require conditions on $\varphi$ and its first derivatives only on arbitrarily small sub-regions of $\mathcal{I}^+\cup\mathcal{I}^-$ surrounding $i^0$ (since we are allowing $u_0, v_0<0$). This agrees with what we would have expected given the results obtained in previous sections. This is because for hyperbolic equations, initial data propagates along the characteristic curves. In this problem, the characteristic curves are the null curves of the metric $g$. Suppose that, having fixed $\varphi$ and its first derivatives to vanish sufficiently quickly on some intervals $\mathcal{I}^+_{u_0}\cup\mathcal{I}^-_{v_0}$, we are unable to extend the solution uniquely into a region $\mathcal{D}^{(u_0,v_0)}_\omega$ for any $\omega>0$. Then for arbitrarily small $\omega>0$, there must exist null curves which propagate from outside either of these intervals into $\mathcal{D}^{(u_0,v_0)}_\omega$. Equivalently, the following condition must fail to hold:\\
\\
\textbf{Null Geodesic Endpoint Condition:} We say that a weakly asymptotically empty and simple spacetime, $(M,g)$, satisfies the null geodesic endpoint condition if, given any $u_0$, $v_0\in\mathbbm{R}$, there exists $\omega>0$ such that any inextendible null geodesic entering $\mathcal{D}^{(u_0,v_0)}_\omega$ must have at least one endpoint in $\mathcal{I}_{v_0}^-$ or $\mathcal{I}^+_{u_0}$.

\begin{figure} 
  \minipage{0.48\textwidth}
  \centering
      \includegraphics[scale=0.3]{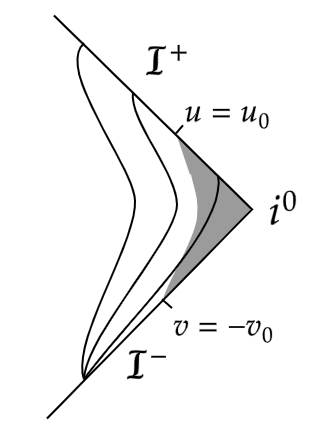}
    \caption{In Schwarzschild spacetime, requiring a null geodesic with past endpoint not in $\mathcal{I}^-_{v_0}$ to enter $\mathcal{D}^{(u_0,v_0)}_\omega$ (shaded region) for arbitrarily small $\omega>0$ forces the future endpoint to lie arbitrarily close to $i^0$.}
    \label{fig:waveeqnschw}
    \endminipage
     \minipage{0.04\textwidth}
      \includegraphics[scale=0.05]{blank.png}
      \endminipage
    \minipage{0.48\textwidth}
 \centering
   \includegraphics[scale=0.3]{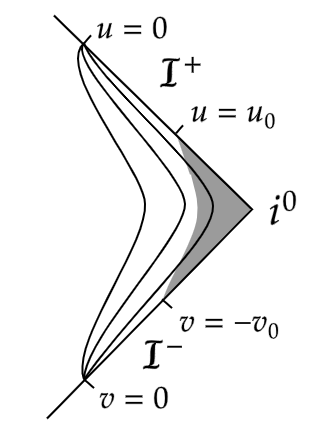}
    \caption{For any $u_0,v_0<0$, null geodesics in Minkowski spacetime with past and future endpoints lying outside $\mathcal{I}^-_{v_0}$ and $\mathcal{I}^+_{u_0}$ (e.g. at $v=0$ and $u=0$) can enter $\mathcal{D}^{(u_0,v_0)}_\omega$ (shaded region) for any $\omega>0$.}
    \label{fig:waveeqnmink}
    \endminipage
\end{figure}

This is the property which Penrose shows \cite{Penrose} holds in positive mass Schwarzschild spacetime in 3+1 dimensions and which we have reviewed in section \ref{Positive Mass Schwarzschild in 3+1 Dimensions}. Indeed for a null geodesic in this spacetime with fixed past endpoint not contained in $\mathcal{I}^-_{v_0}$, the requirement of entering $\mathcal{D}_\omega^{(u_0,v_0)}$ for arbitrarily small $\omega$ is equivalent to letting $R \rightarrow \infty$. This results in the future endpoint sliding along $\mathcal{I}^+$ towards $i^0$ and eventually entering $\mathcal{I}^+_{u_0}$ (see Figure \ref{fig:waveeqnschw}). This is a consequence of the stationarity of the spacetime, which allows us to extend the endpoint results of \cite{Penrose} and section \ref{Positive Mass Schwarzschild in 3+1 Dimensions}. The null geodesics considered there were assumed to satisfy $r=R$ at $t=0$ and hence were symmetric about $t\mapsto -t$. Translations in $t$ will have the effect of moving the endpoints of these null geodesics along null infinity, so we can choose to keep the past endpoint fixed as we let $R\rightarrow\infty$, which we then see results in the future endpoint approaching $i^0$.

For Minkowski spacetime, this endpoint condition does not hold. We can, for example, fix the endpoints to lie at $u=0$ and $v=0$ while choosing the impact parameter to be arbitrarily large, so for any $u_0,v_0<0$, the null geodesic enters $\mathcal{D}_\omega^{(u_0,v_0)}$ for $\omega$ arbitrarily small (see Figure \ref{fig:waveeqnmink}). This is exactly why it is necessary to impose conditions on the solution on
$\mathcal{I}^+_\epsilon\cup\mathcal{I}^-_\epsilon$
for some $\epsilon>0$ in order to achieve the unique continuation result. The solution will vanish in regions $\mathcal{D}^{(\epsilon,\epsilon)}_\omega$ which cannot be entered by any inextendible null geodesic with both endpoints outside $\mathcal{I}^+_\epsilon\cup\mathcal{I}^-_\epsilon$.
\subsection{General Static, Spherically Symmetric Spacetimes}
We saw in section \ref{Positive Mass Schwarzschild in Higher Dimensions} that Penrose's method of proof does not extend to positive mass Schwarzschild in higher dimensions. This is because the null geodesic endpoint condition is no longer satisfied. In particular, if we impose $r=R$ at $t=0$, then at the past and future endpoints we have $v\rightarrow0^-$ and $u\rightarrow0^+$ respectively as we let $R\rightarrow0$. As noted in \cite{LinearWaves}, this is why it is necessary to impose conditions on $\varphi$ on $\mathcal{I}^+_\epsilon\cup\mathcal{I}^-_\epsilon$ for some $\epsilon>0$ (as in Minkowski) in order to obtain a unique continuation result.

As we have seen, to prove that the Penrose property holds for some spacetime we could try to use the same construction as Penrose in \cite{Penrose}. If the metric is stationary then, as discussed in the previous section, showing that non-radial null geodesic endpoints approach $i^0$ as we let $R\rightarrow\infty$ is equivalent to proving that the null geodesic endpoint condition holds. If we now restrict attention further and consider only static, spherically symmetric spacetimes in $d+1$ dimensions ($d\geq3$) then the other result we require for the Penrose construction is that $|\Delta\phi|\geq\pi$ along all non-radial null geodesics (restricted to large $r$ if necessary). A general static, spherically symmetric metric is characterised by two functions, $A(r)$ and $B(r)$, and in $d+1$ dimensions can be written as
\begin{equation}\label{eqn:StaticSphericallySymmetric}
    ds^2=A(r)^2dt^2-B(r)^2dr^2-r^2d\omega_{d-1}^2
\end{equation}
Since both the Penrose property and the null geodesic endpoint condition are conformally invariant, we are only interested in the conformal class of this metric. This allows us to remove one of these functional degrees of freedom. If we conformally re-scale the metric by a factor $A(r)^{-2}$, define a new radial co-ordinate $\tilde{r}=\frac{r}{A(r)}$ and then define the function\footnote{We assume that we can invert the function $\tilde{r}(r)$ (restricted to large $r$ if necessary). This is guaranteed to be the case if we assume, consistent with asymptotic flatness, that $A(r)$ can be expanded as a Taylor series in $1/r$ (as we assume for $C(r)$ in equation (\ref{eqn:expansion})). This ensures that $\tilde{r}(r)$ is a monotonic function (for sufficiently large $r$), and hence is invertible by the inverse function theorem. Recall that we can restrict attention to large $r$ if necessary since the Penrose property is a property near $i^0$.} $C(\tilde{r})=\frac{B(r(\tilde{r}))}{A(r(\tilde{r}))}\frac{dr}{d\tilde{r}
}$, it suffices to consider a metric of the form (dropping tildes):
\begin{equation}\label{eqn:SSSS}
    \begin{split}
        ds^2=dt^2-C(r)^2dr^2-r^2d\omega_{d-1}^2
    \end{split}
\end{equation}
Using Lemma \ref{spherical symmetry}, we can restrict attention to the $\Sigma\times S^1$ submanifold with co-ordinates $(t,r,\phi)$, on which the induced metric is 
\begin{equation}\label{eqn:induced}
    ds^2=dt^2-C(r)^2dr^2-r^2d\phi^2
\end{equation}
Consistent with asymptotic flatness, we assume we can expand:
\begin{equation}\label{eqn:expansion}
    C(r)=1+\frac{C_{d-2}}{r^{d-2}}+O\left(\frac{1}{r^{d-3}}\right)
\end{equation}
A calculation similar to the one carried out in \cite{Penrose} shows that the null geodesic endpoint condition is satisfied if and only if $C_{d-2}>0$ and $d=3$. This is precisely the condition required to get the logarithmic divergence observed by Penrose.  On the other hand, the angular condition ($|\Delta\phi|\geq\pi$ along non-radial null geodesics restricted to sufficiently large $r$) is satisfied if and only if $C(r)>1$ for $r$ sufficiently large. We therefore observe that this is a weaker condition than the null geodesic endpoint condition and hence that Penrose's construction works if and only if $C_{d-2}>0$ and $d=3$.

To determine if the Penrose property is satisfied for a metric of the form (\ref{eqn:SSSS}) we can use the same arguments as were used in previous sections for Schwarzschild spacetime. We have seen that if $d=3$ and $C_1>0$ then the Penrose construction suffices to show that the Penrose property is satisfied. If $d=3$ and $C_1<0$ or $d>3$ and $C_{d-2}>0$, then the comparison arguments outlined in Sections \ref{Negative Mass Schwarzschild in 3+1 Dimensions} and \ref{Positive Mass Schwarzschild in Higher Dimensions} still hold and show that the Penrose property is not satisfied. If $C_{d-2}<0$ and $d>3$, we can use the same arguments as section \ref{Negative Mass Schwarzschild in Higher Dimensions} to again show that the Penrose property is not satisfied. The only remaining case is $C_{d-2}=0$ ($d\geq3$). This case is straightforward because either $C(r)\equiv1$, in which case the metric is flat and we have seen that the Penrose property does not hold, or we have
$$C(r)=1+\frac{C_{d'-2}}{r^{d'-2}}+O\left(\frac{1}{r^{d'-3}}\right)$$
for some $d'>d\geq3$, where $C_{d'-2}\neq0$. In this case, the problem reduces to the question of whether the Penrose property is satisfied by some $d'+1$ dimensional spacetime with metric of the form (\ref{eqn:SSSS}) (and inducing a metric of the form (\ref{eqn:induced})), where $C_{d'-1}\neq0$. But we have seen that the Penrose property is never satisfied for such higher dimensional spacetimes, so we conclude that for static, spherically symmetric spacetimes in $d+1$ dimensions ($d\geq3$), we have
\begin{center}
\boxed{\text{Penrose property}\iff\left[ d=3\text{ and }C_1>0\right]\iff\text{Null Geodesic Endpoint Condition}}
\end{center}
where $C_1$ is the constant arising from considering the conformally related metric of the form (\ref{eqn:SSSS}) and expanding $C(r)$ as in equation (\ref{eqn:expansion}).

The constant $C_{d-2}$ is equal to the ADM mass of the spacetime. However it does not make sense to phrase the Penrose property in terms of the ADM mass. This is because, unlike the Penrose property, the ADM mass (and even its sign) is not a conformally invariant quantity. In particular, in $3+1$ dimensions it is possible to have two metrics of the form (\ref{eqn:StaticSphericallySymmetric}) with the same functions $B(r)$ (and hence the same ADM mass) where one satisfies the Penrose property but the other does not (we can choose the function $A(r)$ such that the constant $C_{d-2}$ has either sign).

\end{document}